\documentclass[%
 reprint,
 amsmath,amssymb,
 aps,
showkeys]{revtex4-2}

\usepackage{graphicx}% Include figure files
\usepackage{dcolumn}% Align table columns on decimal point
\usepackage{bm}% bold math
\usepackage{array,supertabular,hhline,enumitem,hyperref}
\usepackage{array}
\usepackage{placeins}
\usepackage{float}
\usepackage{multirow}
\usepackage{makecell}
\usepackage{titlesec}
\usepackage{setspace}
\usepackage{amsmath}

\titlespacing*{\section}{0pt}{2ex}{2ex}
\begin{document}

%\preprint{APS/123-QED}

\title{Planar Nanofluidic Memristors Enabled by Surface Charge Gradient}% Force line breaks with \\

\author{Zhongyuan Zhao}
\author{Ziyi Yin}
\author{Chudi Qi}
\author{Yuheng Li}
\author{Shoushan Fan}
\author{Qunqing Li}
\author{Yang Wei}
 \email{weiyang@tsinghua.edu.cn}
 \affiliation{Department of Physics and
 	Tsinghua-Foxconn Nanotechnology Research Center, State Key Laboratory of Low-Dimensional Quantum Physics, Tsinghua University, Beijing 100084, China.}


\begin{abstract}
Nanofluidic memristors, exploiting ion transport in nanochannels, hold promise for neuromorphic applications. A planar architecture is particularly desired for scalable integration with established micro- and nanofabrication technologies. Here, using the Poisson–Nernst–Planck framework, we theoretically propose planar nanofluidic memristors enabled by surface charge gradient, providing an alternative to the commonly used geometrically asymmetric architectures. Memristive behaviors are governed by a diffusion-mediated ``secondary enrichment effect''. By systematically solving the PNP equations, we obtain the scaling of the characteristic memory time across the parameter space.  We also reveal that the memory effect is related to the first-order moment of surface charge, for arbitrary charge profiles. These results provide a theoretical basis for rationally designing and optimizing planar nanofluidic memristors through spatially patterned surface charge.
%\begin{description}
%\end{description}
\end{abstract}

%[showkeys]%Use showkeys class option if keyword
\maketitle

%\tableofcontents
%\onecolumngrid  
\section{Introduction}

	Current computing architectures separate computation and memory, facing ever-growing memory-wall bottlenecks \cite{078,082}. Memristors, in contrast, integrate information storage and processing in a single circuit element, thus holding great promise for in-memory and neuromorphic computing \cite{075,057}. What defines a memristor is a tunable internal state, which encodes the history of applied stimuli and accordingly determines the output characteristics \cite{075,087}. In conventional electron- and hole-based memristive systems, the internal states arise from various mechanisms, including the redistribution of oxygen vacancies, charge trap occupation, ferroelectric polarization and the phase transitions \cite{080,079,086}. Nanofluidic devices, which facilitate ion transport in nano-confined channels, open up a unique avenue for memristor-based applications \cite{006,054,065,084,081}. Due to the confinement effect, the surface charges on the channel walls considerably modulate ion concentrations inside the channel, leading to a tunable ionic conductance \cite{010,008,074}. The ion concentration can serve as the internal state and the tunable conductance makes nanofluidic memristors feasible. Beyond the role as charge carriers, ions possess extra chemical information and can participate in diverse physicochemical interactions, offering nanofluidic memristors functionalities unavailable to their electronic counterparts \cite{083,009,031}. Moreover, the operation in aqueous environments provides intrinsic compatibilities with biological systems, making nanofluidic memristors particularly promising for neuromorphic computing and emerging technologies such as brain-machine interfaces \cite{081,083,076}.
	
	Early studies prototyped nanofluidic memristors with geometrically asymmetric nanochannels, particularly conical nanopores \cite{088,089,090}. Such devices have exhibited conductance potentiation/depression, learning/forgetting and other forms of synaptic plasticity \cite{063,076,059}. Subsequent theoretical studies and experiments have further demonstrated that conical devices can reproduce neuronal action potentials and trains of action potentials, highlighting their potential as ionic building blocks for brain-inspired computing \cite{077,060,061}. Despite these advances, the geometrically asymmetric configuration and specialized fabrication of conical nanopores pose challenges for scalable integration via established planar micro- and nanofabrication technologies. Although multipore membrane has been successfully fabricated \cite{091,092,093}, achieving densely and systematically integrated device arrays remains a significant challenge. This is particularly vital in constructing neuromorphic systems, since neuronal functions emerge from the coordinated functioning of multiple ionic channels. A planar architecture compatible with mature planar fabrication technologies could therefore provide an alternative route toward scalable integration of ionic elements. Notably, a rational design strategy for planar memristors can be demonstrated theoretically prior to experimental realization.
	\begin{figure*}[htbp]
		\includegraphics[width=0.89\textwidth]{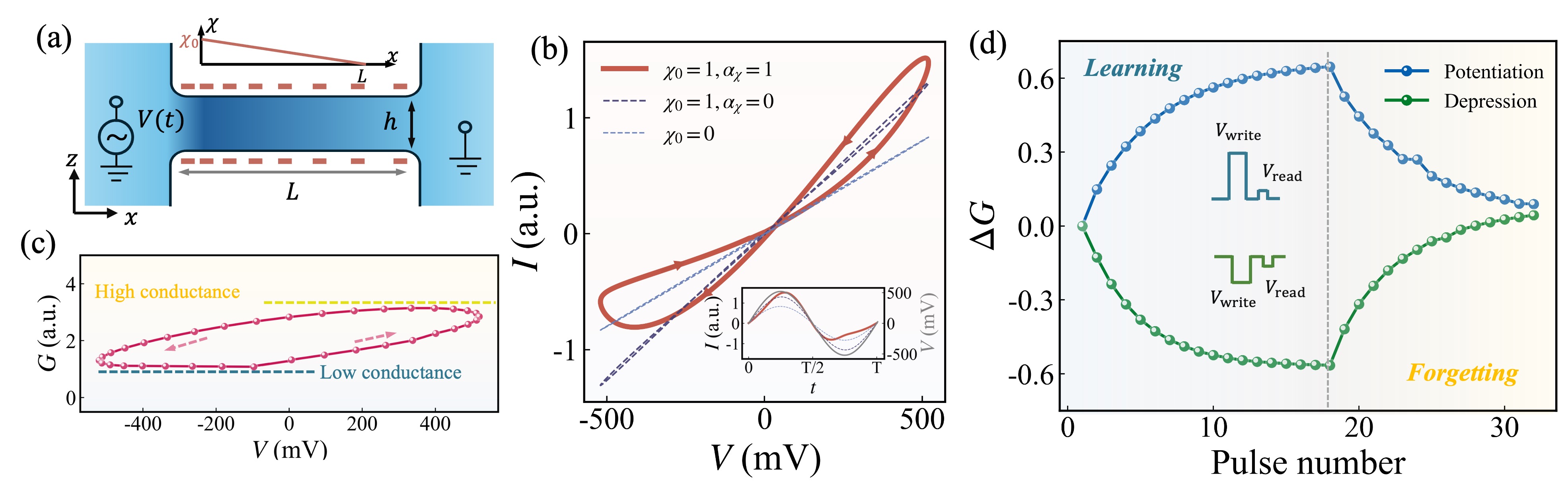}% Here is how to import EPS art
		\caption{(a) Schematic of the planar nanochannel. (b) Simulated current–voltage characteristics for an uncharged channel ($\chi_0=0$), a uniform surface charge ($\chi_0=1, \alpha_\chi=0$) and a surface charge gradient ($\chi_0=1, \alpha_\chi=1$). (c) $G -V$ characteristics, showing hysteretic switching between low- and high-conductance states. (d) Synaptic behaviors of the memristor.}
		\label{fig:Figure 1}
	\end{figure*}
	
	Ion dynamics can be effectively described by the Poisson-Nernst-Planck (PNP) framework, which is believed to apply down to 1 nm and should cover a wide range of nanofluidic devices \cite{006,054}. At equilibrium, the Nernst-Planck equation yields Boltzmann distribution and the Poisson-Boltzmann (PB) equation is easier to solve. Our previous work on PB theory states that, the ubiquitous surface charge induces an additional ionic conductance as a function of $\chi/\gamma$, in which $\gamma$ is the relative channel height and $\chi$ is proportional to the surface charge density \cite{074}. A time-varying voltage across the channel drives the system from equilibrium into a dynamic state, where history-dependent conductance can occur. Experimental studies have shown that a conical geometry, which in essence brings a gradient in $\gamma$, can lead to memristive behaviors. Then from an intuitive argument, a gradient in surface charge $\chi$ may have the same effect. In this work, we numerically solve the PNP equations via customized finite difference methods (FDM). We confirm that in a planar nanochannel, introducing surface charge gradient can indeed bring robust memristive behaviors, which is attributed to a ``secondary enrichment effect''. By systematically solving the PNP equations, we reveal the scaling behaviors of the characteristic time of ionic memory over the parameter space, which provides direct guidance for experimental design and multiscale modelling. We further generalize to various surface charge configurations $\chi(x)$ and uncover a linear relationship between the memory effect and the first-order moment of  $\chi$. Lastly, as the strategy of introducing gradient surface properties has been widely applied in distinct systems \cite{068,067,072}, we discuss plausible routes to experimentally realize planar nanofluidic memristors with patterned surface charge and their optimization principles.
\section{Model, phenomenon and mechanism}
	\begin{figure*}[htbp!!!!!!!]
		\includegraphics[width=0.89\textwidth]{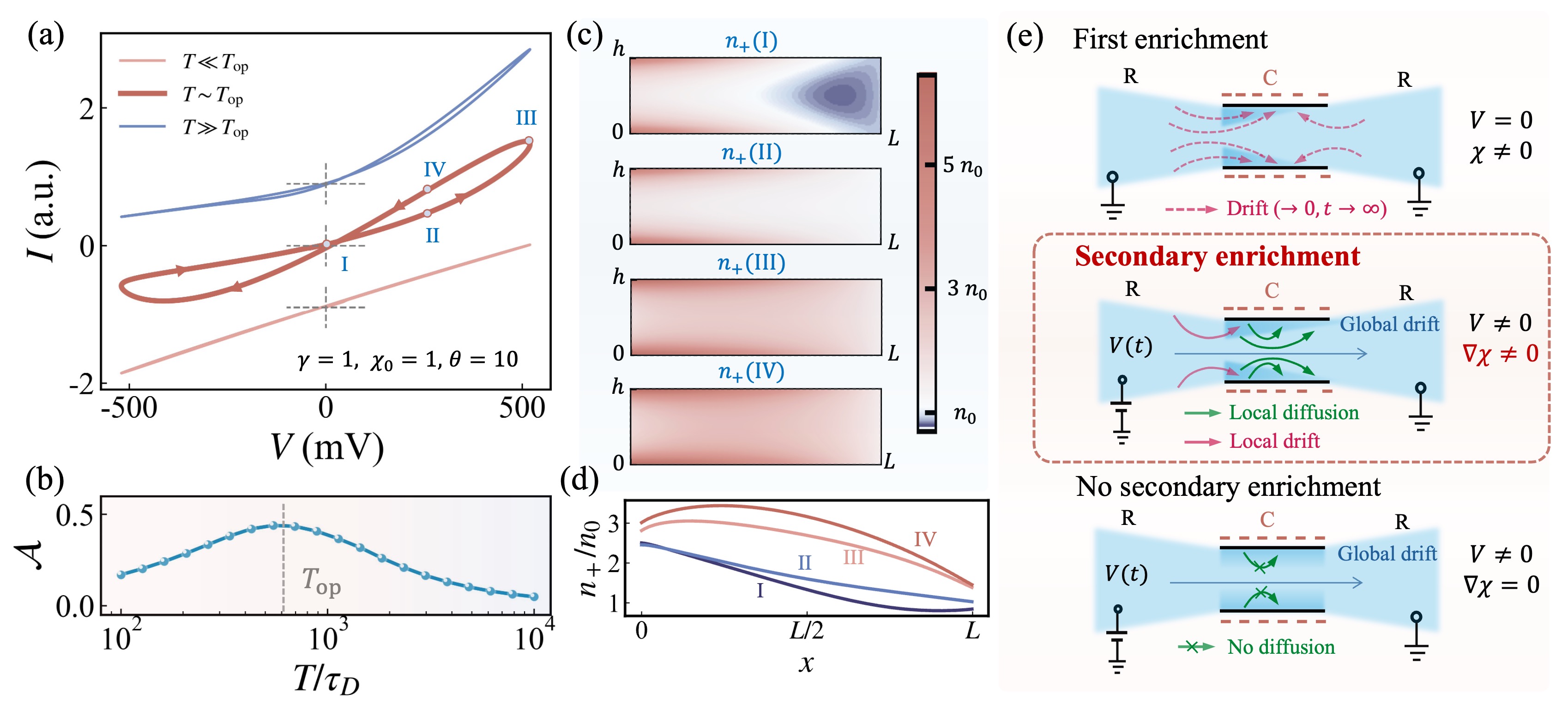}% Here is how to import EPS art
		\caption{(a) The frequency-dependent $I$--$V$ loops. (b) The hysteresis area as a function of the period of $V(t)$ (c) Snapshots of cation distributions. (d) The cation density profile $n_+(x)=\frac{1}{h}\int_0^h n_+(x,z)\mathrm{d}z$. (e) Illustration of first and secondary enrichments. R, C are short for reservoir and channel, respectively. For the first enrichment, the drift process is denoted by a dashed line, indicating that the drift will cease when equilibrium is reached and electric double layers are formed. While for the secondary enrichment, the drift and diffusion processes are denoted by solid lines, indicative of a continuous nature as long as $V(t)\neq0$.}
		\label{fig:Figure 2}
	\end{figure*}
	To study the surface-charge-modulated ion dynamics, a reservoir-channel-reservoir model is established, as depicted by Fig. \ref{fig:Figure 1}(a). The channel is a nano-slit with length $L$ and height $h$. A voltage $V(t)$ is applied across the channel to the reservoirs with a constant ion concentration $n_0$. Surface charges exist on both walls with a spatially varying density $\Sigma(x)$. According to the Nernst-Planck-Poisson framework, the ionic fluxes are:
	\begin{equation}
		\mathbf{J}_\alpha
		=
		- D \left(
		\nabla n_\alpha
		+ \frac{z_\alpha e}{k_\mathrm{B}T} n_\alpha \nabla \varphi
		\right),
		\qquad \alpha\in\{+,-\},
		\label{eq:NP euqation main}
	\end{equation}
	in which $D$ is the diffusion coefficient and $\varphi$ is the electric potential satisfying Poisson equation:
		\begin{equation}
		\nabla^2 \varphi
		=
		-\frac{e}{\varepsilon}
		\left(
		z_+ n_+ + z_- n_-
		\right).
		\label{eq:Poisson euqation main}
	\end{equation}
	Unless otherwise specified, $z_\pm=\pm1$. With the continuity equations below, eqs. \ref{eq:NP euqation main}-\ref{eq:continuity euqation main} constitute a set of self-consistent equations and the ion dynamics can be resolved.
	\begin{equation}
		\frac{\partial n_\alpha}{\partial t}
		+
		\nabla\cdot\mathbf{J}_\alpha
		=
		0.
		\label{eq:continuity euqation main}
	\end{equation}
	By introducing the Debye length $l_\mathrm{D}=\sqrt{\varepsilon k_\mathrm{B} T/2 e^2 n_0}$ and a time scale $\tau_\mathrm{D}=l_\mathrm{D}^2/D$, the PNP equations can be nondimensionalized (detailed in Supplementary Material, SM, section 1), and three key dimensionless parameters emerge, following the definitions in \cite{074}:
	\begin{equation}
		\gamma=h/4l_\mathrm{D},\quad \theta=L/h, \quad\chi=\frac{e^2l_\mathrm{D}}{2\varepsilon k_\mathrm{B}T}\Sigma,
	\end{equation}
	which are the relative channel height, relative length and relative charge density, respectively. The PNP equations can be solved with respect to $\gamma$, $\theta$, $\chi(x)$ and $V(t)$, via a customized finite difference method (SM, section 2).

	In Fig. \ref{fig:Figure 1}(b), the reduced ionic currents $I=\frac{1}{h}\int_0^h (\bar{J}_{x,+}-\bar{J}_{x,-})|_{x=0}\mathrm{d}z$ as functions of $V(t)=V_{\mathrm{max}}\sin(2\pi t/T)$ are calculated for $\gamma=1, \theta=10, T=600 \tau_\mathrm{D}$ and at different surface charge configurations $\chi(x)=\chi_0 (1-\alpha_\chi \frac{x}{L})$.
	In the absence of surface charge ($\chi_0=0$), the ionic transport is purely bulk-type and ohmic (the light blue dashed line in Fig. \ref{fig:Figure 1}(b)). In the presence of surface charge ($\chi_0=1$), the channel experiences a ``first enrichment'' (illustrated by the first panel of Fig. \ref{fig:Figure 2}(e)). Ions of opposite polarity to the surface charge (counter-ions) are enriched until an equilibrium state, leading to increased ion concentration and flux \cite{074}:
	\begin{equation}
		n(x), J_x(x)\propto 2n_0 \sqrt{1+(\frac{\chi(x)}{\gamma})^2}.
	\end{equation} 
	The dark blue dashed line in Fig. \ref{fig:Figure 1}(b) suggests the enhancement of ionic conductance at uniform  distributed surface charges ($\alpha_\chi=0$). No hysteresis appears until a surface charge gradient ($\alpha_\chi=1$) is introduced, as signified by the red line in Fig. \ref{fig:Figure 1}(b). The non-linear pinched $I$--$V$ loop is the hallmark of a memristor \cite{075}, and it can be recognized in Fig. \ref{fig:Figure 1}(c) that the channel shifts from a low conductance ($G=I/V$) state to a high conductance state at positive $V(t)$ and returns to the low conductance state at negative $V(t)$ hysteretically. The ionic state can be further modulated by voltage pulses, as shown in Fig. \ref{fig:Figure 1}(d): repeated write pulses induce conductance potentiation or depression, while the conductance gradually relaxes after the write pulses are removed, demonstrating learning and forgetting functionalities. These results confirm that by introducing $\nabla \chi$, ionic memristors can be achieved and utilized in neuromorphic computing.
	
	Next we begin rationalizing the underlying physics. In a fluidic medium, ion transport is in principle entangled with hydrodynamic effects (electro-osmosis). Previous analyses suggest hydrodynamic contributions become significant only when $\gamma\chi\gg1$ and $\chi/\gamma\gg1$ \cite{074}. Thus in the intermediate parameter range ($\chi\sim1$ and $\gamma\sim1$) considered in this work, we neglect the fluidic convection and retain only the ionic diffusion (driven by concentration gradient) and drift (driven by electric fields) as in eq. \ref{eq:NP euqation main}. By isolating the essential processes, a more transparent physical origin can be found. Define $\mathcal{A}$ to quantify the area formed by the hysteretic $I$--$V$ loop \cite{059}:
	\begin{equation}
		\mathcal{A}=-\frac{4}{\Delta I\Delta V}\oint |I(t)|\mathrm{d}V.
	\end{equation} 
	For the applied sinusoidal voltage, an optimal period $T_{\mathrm{op}}$ exists where $\mathcal{A}$ reaches maximum, indicated by the peak on the blue line in Fig. \ref{fig:Figure 2}(b). $\mathcal{A}\rightarrow0$ either when $T\gg T_{\mathrm{op}}$ or $T\ll T_{\mathrm{op}}$ as shown in Fig. \ref{fig:Figure 2}(a). The frequency dependence is another signature of a memristor and the characteristic memory time $T_{\mathrm{op}}$ should reflect the memristive mechanism. The drift process has a definite time scale of $\tau_\mathrm{D}=l_\mathrm{D}^2/D=\varepsilon/\sigma_0$, in which $\sigma_0$ is the bulk conductivity (SM, section 3 and Fig. S5). Whereas diffusion has a time scale of $\tau_\mathrm{di}=l_\mathrm{di}^2/D$, depending on the diffusion length $l_\mathrm{di}$. According to Fig. \ref{fig:Figure 2}(b), $T_{\mathrm{op}}$ is much larger than the drift time $\tau_\mathrm{D}$, so diffusion must have participated in and even dominated the memristive response.
	
	The roles of diffusion and drift are unraveled by firstly examining the evolution of ion distribution over $T_\mathrm{op}$. With four time-stamps (I - IV) marked in Fig. \ref{fig:Figure 2}(a), the respective mappings of cation concentration $n_+(x,z)$ are stacked in Fig. \ref{fig:Figure 2}(c), and the line-profiles are shown in Fig. \ref{fig:Figure 2}(d). Here we assume negative surface charges and thus cations are the major carriers (equally effective for positive $\Sigma$ except for a backward voltage $V\rightarrow-V$ and anions as the major carriers). By applying a positive $V(t)$, a secondary enrichment of cations happens, starting at the high-concentration side (I) and extending to the entire channel (IV). The further enriched cations contribute to an increased current. Conversely, a negative $V(t)$ cancels the enrichment and reduces $|I|$. 
	\begin{figure*}[htbp!!!!!]
		\includegraphics[width=0.862\textwidth]{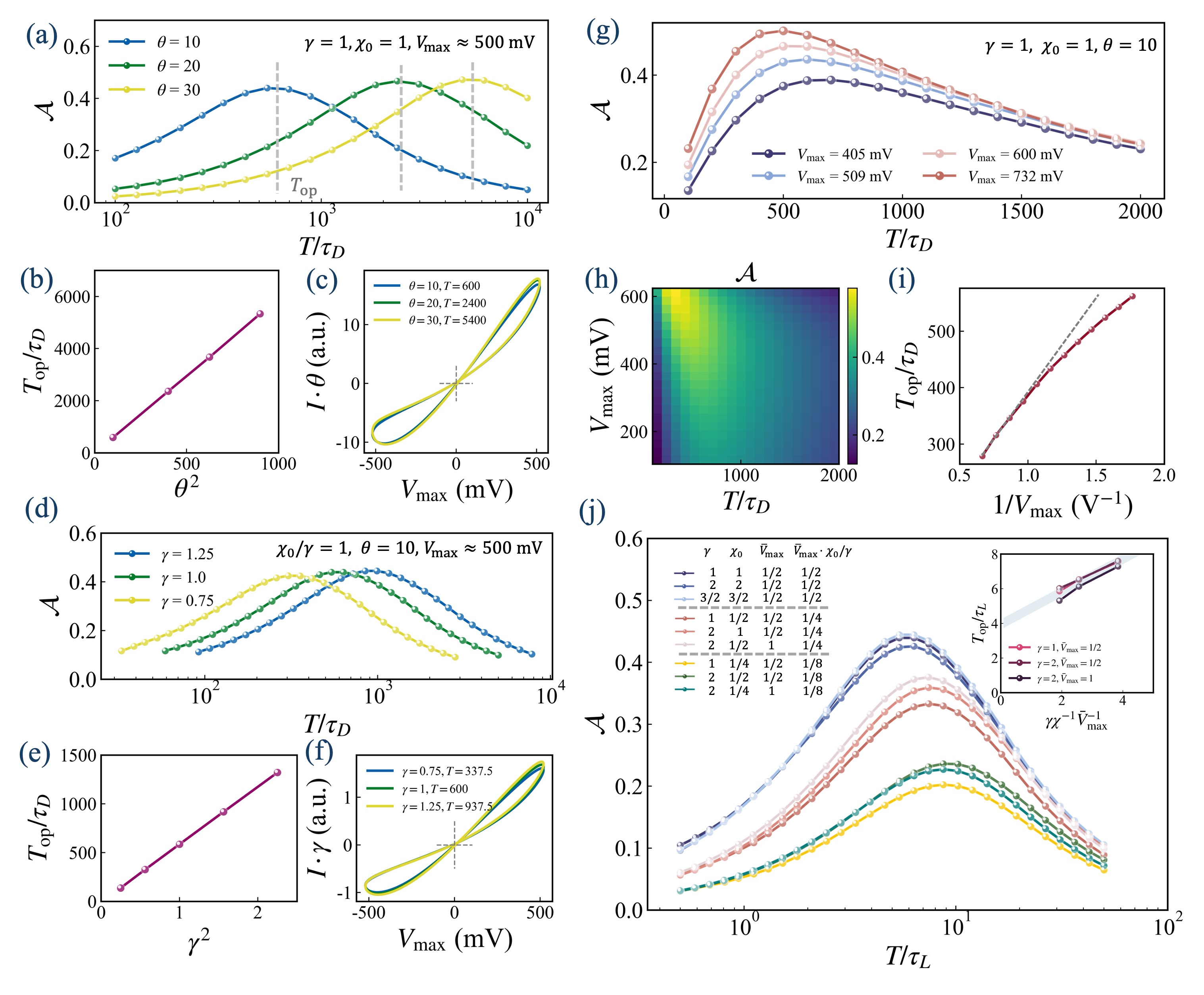}% Here is how to import EPS art
		\caption{Memristive characteristics over the entire parameter space. (a) $\mathcal{A}(T)$ at different $\theta$. (b) The linear relationship between $T_\mathrm{op}$ and $\theta^2$. (c) The $I$--$V$ loops at different $\theta$. (d) $\mathcal{A}(T)$ at different $\gamma$. (e) The linear relationship between $T_\mathrm{op}$ and $\gamma^2$. (f) The $I$--$V$ loops at different $\gamma$.  (g) $\mathcal{A}(T)$ at different $V_\mathrm{max}$. (h) The mapping of $\mathcal{A}(T,V_\mathrm{max})$. (i) $T_\mathrm{op}$ as a function of $1/V_\mathrm{max}$. (j) $T_\mathrm{op}$ at different combinations of $\gamma$, $\chi_0$, $V_\mathrm{max}$. The inset shows $T_\mathrm{op}$ as a function of the product of  $\gamma$, $\chi_0^{-1}$, $V_\mathrm{max}^{-1}$.}
		\label{fig:Figure 3}
	\end{figure*}
	
	To illustrate this secondary enrichment effect, we take a look at the zero-voltage state (I). Apparently a concentration gradient exists(the first panel of Fig. \ref{fig:Figure 2}(c)), which surely brings a tendency for diffusion. Meanwhile the non-uniform surface charge also produces an electric field in the -x direction (SM, Figure S7), bringing a tendency for drift. At equilibrium ($V=0, t\rightarrow\infty$), the tendencies for diffusion and drift are perfectly balanced (reminiscent of a PN junction), yielding zero net ionic fluxes $J_\pm$ at any place (detailed balance). However, by exerting $V$, the external electric field not only brings a global drift tendency which generates a macroscopic current $I\propto V/L$, but also locally disrupts the detailed balance between drift and diffusion. The diffusion-mediated mechanism is illustrated in Fig. \ref{fig:Figure 2}(e). Via diffusion, the ions move from the high-concentration (HC) zone to the low-concentration (LC) zone (secondary enrichment), increasing $n$ at the LC zone. And simultaneously, the HC zone is replenished by the reservoir via drift (first enrichment), which is much faster ($\sim \tau_\mathrm{D}$, SM, Fig. S5), and thus maintains high-concentration. Henceforward, under the combined effects of both diffusion and drift, the high-concentration profile is ``replicated'' from the HC zone to the entire channel, as evidenced by Fig. \ref{fig:Figure 2}(d). The formation and relaxation (Fig. \ref{fig:Figure 2}(d)) of this non-equilibrium state are constrained by the slower diffusion and thus have a diffusive time scale $\tau_\mathrm{di}\gg \tau_\mathrm{D}$. 

		\begin{table*}[]
		\centering
		\renewcommand{\arraystretch}{1.2}
		\setlength{\tabcolsep}{12.0pt}
		\begin{tabular}{c|c|c}
			\hline
			
			\textbf{Equilibrium} &
			\textbf{Steady state} &
			\textbf{Non-equilibrium}
			\\
			\hline
			$\vec{J}=0$&{$\vec{J}\neq0$}&{$\vec{J}\neq0$}\\
			$\nabla\cdot\vec{J}=0$&$\nabla\cdot\vec{J}=0$&{$\nabla\cdot\vec{J}\neq0$}\\
			$\displaystyle \partial_t\vec{J}=0$&
			$\displaystyle \partial_t\vec{J}=0$&
			{$\displaystyle\partial_t\vec{J}\neq0$}
			\\
			\hline
			\begin{tabular}{c}
				Boltzmann\\distribution		
			\end{tabular}
			&
			\begin{tabular}{c}
				Quasi-Boltzmann\\distribution
			\end{tabular}
			&
			\begin{tabular}{c}
				Non-Boltzmann\\distribution
			\end{tabular}
			\\
			\hline
			\multicolumn{2}{c|}{\begin{tabular}{c}
				Ohmic behavior  \cite{074}, $	I\propto\sqrt{1+\left(\frac{\chi}{\gamma}\right)^2}\cdot V$ 
			\end{tabular}}
			%\vspace{0.4em}
			&
			\begin{tabular}{c}
				Memory effect,
				$I\propto\int_{0}^{t}V(\tau)G_{\mathrm{mk}}(t-\tau)\,d\tau$
			\end{tabular}
			\\
			\hline
		\end{tabular}
		\caption{The formalism summerizing equilibrium, steady state and non-equilibrium state of ionic dynamics.}
		\label{table: equilibrium to non-equilibrium}
	\end{table*}
	\begin{figure*}[htbp]
		\includegraphics[width=0.84\textwidth]{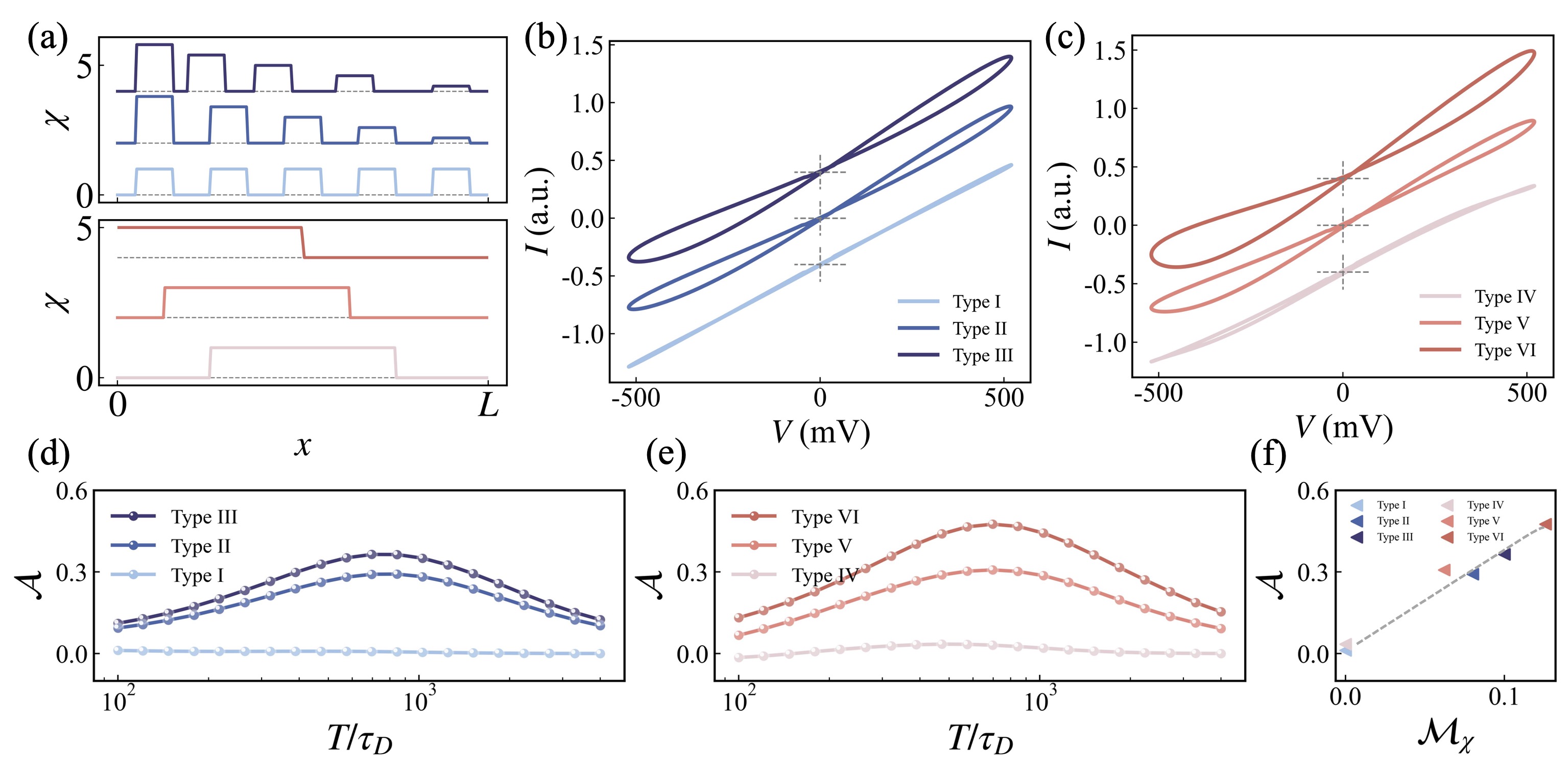}% Here is how to import EPS art
		\caption{Generalization to arbitrary surface-charge configurations. 
			(a) Representative spatial distributions of the surface charge $\chi(x)$. 
			(b,c) Corresponding $I$--$V$ characteristics for the different surface charge configurations. (d,e) $\mathcal{A}(T)$ for the corresponding surface-charge configurations. 
			(f) Maximum hysteresis area $\mathcal{A}_{\max}$ as a function of the first spatial moment of the surface charge.}
		\label{fig:Figure 4}
	\end{figure*}

%\clearpage 
\section{Dependence of characteristic time}
	The above arguments provide an explicit understanding of the memristive mechanism. Further, a concrete knowledge of memristive performances at any device and operation parameters is required.  In this section we obatin the finer dependence of $T_\mathrm{op}$ on parameters $\gamma$, $\chi_0$, $\theta$, and $V_\mathrm{max}$, by systematically solving the PNP equations and analyzing the hysteresis characteristics $\mathcal{A}(T)$. These analyses not only help experimentally match optimal operating conditions to diverse device configurations, but also enable building lightweight  ralaxation-time approximation models with an accurate $T_\mathrm{op}$, which is a practical strategy for multiscale modelling.
	
	We therefore examine the characteristic operating time $T_\mathrm{op}$ separately in terms of extensive ($\theta$ and $\gamma$) and intensive quantities ($V_\mathrm{max}$ and $\chi_0/\gamma$), establishing scaling relations with the former and resolving its finer dependence on the latter. As the extensive parameters are varied in Figs.~\ref{fig:Figure 3}(a)-(f), the $\mathcal{A}(T)$ spectra exhibit an essentially identical profile with a systematic shift along the $T/\tau_D$ axis. The extracted $T_\mathrm{op}$ follows a quadratic scaling with both $\theta$ and $\gamma$, i.e., $T_{\mathrm{op}}\propto\theta^2$ and $T_{\mathrm{op}}\propto\gamma^2$, as shown in Figs.~\ref{fig:Figure 3}(b) and \ref{fig:Figure 3}(e). Hence $T_\mathrm{op}\propto\tau_\mathrm{L}\equiv\gamma^2\theta^2\tau_\mathrm{D}=L^2/16D$, again supporting the preceding argument of the diffusion-mediated enrichment over the entire channel. Moreover, when the corresponding time and current scales are properly normalized, the $I$--$V$ characteristics at different values of $\theta$ and $\gamma$ collapse onto nearly identical curves (Figs.~\ref{fig:Figure 3}(c) and \ref{fig:Figure 3}(f)), demonstrating the self-similar memristive response with respect to the extensive quantities. 
		
	Next we examine the finer dependence $T_\mathrm{op}=f(V_\mathrm{max},\chi_0/\gamma)\cdot\tau_\mathrm{L}$. For a fixed device configuration, increasing $V_{\mathrm{max}}$ shifts the $\mathcal{A}(T)$ spectrum toward shorter operation times, while also modifying the maximum hysteresis area, as shown in Fig.~\ref{fig:Figure 3}(g). Figure~\ref{fig:Figure 3}(h) summarizes $\mathcal{A}$ in the $T$--$V_{\mathrm{max}}$ parameter space, where the ridge of maximum $\mathcal{A}$ traces the optimal operating condition. The extracted $T_{\mathrm{op}}/\tau_{\mathrm{D}}$ decreases with $V_{\mathrm{max}}$ (Fig.~\ref{fig:Figure 3}(i)), demonstrating that the driving amplitude provides an additiona control over the operating time. Finally, we seek a unified description of the finer dependence by simultaneously varying $V_{\mathrm{max}}$, $\chi_0$, and $\gamma$ (Fig.~\ref{fig:Figure 3}(j)). Remarkably, the resulting $\mathcal{A}(T)$ spectra can be grouped into distinct families according to the combined parameter $\bar{V}_{\mathrm{max}}\chi_0/\gamma$ ($\bar{V}_{\mathrm{max}}= V_{\mathrm{max}}/V_0, V_0=1~\mathrm{V}$): cases with different individual values of $\bar{V}_{\mathrm{max}}$, $\chi_0$, and $\gamma$ exhibit nearly identical spectral profiles when this combination is comparable. The inset of Fig.~\ref{fig:Figure 3}(j) further shows that $T_{\mathrm{op}}$ for different spectral families follows a unified dependence on this combined parameter. Within the parameter range investigated in this work (more evidence in SM, section 3.3), we obtain a useful full expression:
	\begin{equation}
		T_{\mathrm{op}}=
		\frac{L^2}{4D}
		\left(
		1+\frac{\gamma}{4\bar{V}_{\mathrm{max}}\chi_0}
		\right).
	\end{equation}
	This relation provides direct guidance for designing device and operating parameters according to the desired memory timescale.
\section{Generalization to arbitrary charge profiles}

	The secondary enrichment effect in the presence of gradient surface charges can be further unified to the mathematical formalism in Table. \ref{table: equilibrium to non-equilibrium}. What distinguishes a non-equilibrium and time-evolving transport from a steady-state transport is whether $\nabla\cdot\vec{J}$ vanishes, enlighted by the structure of PNP equations. This is because $\nabla\cdot\vec{J}=0$ gives static ionic distributions ($\partial_t n=0$ from eq.\ref{eq:continuity euqation main}), leads to static electric potentials and results in static ionic fluxes. In contrast $\nabla\cdot\vec{J}\neq0$ induces ionic redistributions ($\partial_t n\neq0$) and results in time-varying fluxes ($\partial_t \vec{J}\neq 0$). In a physical sense, factors that introduce spatial inhomogeneities to the ionic fluxes ($J\propto\sqrt{1+\left({\chi}/{\gamma}\right)^2}$) will probably result in memory effects. Such factors include the conical geometry ($\nabla\gamma$) and the herein proposed surface charge gradient ($\nabla\chi$ from $\chi(x)=\chi_0(1-x/L)$). And the case may be generalized to arbitrary surface charge configuration $\chi(x)$ with $\nabla\chi\neq 0$.
	
	We consider a series of nonuniform surface-charge configurations $\chi(x)$ beyond the linear gradient in Fig.~\ref{fig:Figure 4}(a). The corresponding $I$--$V$ characteristics exhibit varying degrees of hysteresis in Figs.~\ref{fig:Figure 4}(b) and (c). Extracting the maximum hysteresis area on each curve in Figs.~\ref{fig:Figure 4}(d) and (e), and defining the first spatial moment of the surface charge:
	\begin{equation}
		\mathcal{M}_\chi=\frac{1}{L}\int_0^L \chi(x)\cdot(\frac{L}{2}-x),
	\end{equation}
	a linear relationship is found as shown in Fig.~\ref{fig:Figure 4}(f). This result demonstrates that a linearly varying surface charge profile is not a prerequisite for realizing memristive behavior. Rather, the memory strength is determined by the spatial distribution of surface charge through its first moment. Such a quantitative descriptor provides an additional design degree of freedom for nanofluidic memristors, allowing the channel conductance and the first moment of surface charge to be co-optimized according to the practical constraints of fabrication techniques and the material systems.
	
	The proposed strategy can be experimentally implemented using established planar nanofluidic fabrication techniques. A surface-charge pattern with a desired $\mathcal{M}_{\chi}$ can be first defined by lithography. Then the surfaces can be activated through plasma/RIE or electron-beam treatment, for planar 2D nanochannels fabricated by van der Waals assembly \cite{015,066} or oxide channels fabricated by sacrificial-layer method \cite{011}. When entering the sub-nanometer realm, this surface-charge degree of freedom may act in combination with effects such as ionic blockade and Bjerrum pairing \cite{049,052,066}, offering opportunities for more diverse memristive behaviors.

	\section{Conclusions}
	
	In summary, we theoretically demonstrate a planar nanofluidic memristor enabled by a surface charge gradient. By solving the Poisson--Nernst--Planck equations, we identify a diffusion-mediated ``secondary enrichment effect'' as the origin of the memristive response. We systematically obtain the dependence of operating time on the driving voltage, surface charge, and channel geometry. Arbitrary surface charge profiles can give memristive characteristics, determined by their first spatial moment. These results provide a practical framework for designing and optimizing planar nanofluidic memristors using spatially patterned surface charges.

\begin{acknowledgments}
	
This work was financially supported by the National Key Research and Development Program of China (Grant No.
2022YFA1203400), the National Natural Science Foundation of China (Grant No. 61774090) and the Key-Area Research and
Development Program of Guangdong Province (Grant No. 2020B010169001).

\end{acknowledgments}
\bibliography{manuscript_v0.bib}% Produces the bibliography via BibTeX.
\appendix  
%\clearpage
\onecolumngrid          % 如果正文是双栏，Supplement通常改成单栏

%             % 可有可无

\setcounter{figure}{0}
\renewcommand{\thefigure}{S\arabic{figure}}

\setcounter{table}{0}
\renewcommand{\thetable}{S\arabic{table}}

\setcounter{equation}{0}
\renewcommand{\theequation}{S\arabic{equation}}

% =========================================
% Supplemental Material
% =========================================

% ================================
% Supplemental Material title page
% ================================
\setcounter{secnumdepth}{2}
\makeatletter

\setcounter{section}{0}
\renewcommand{\thesection}{\arabic{section}}
\renewcommand{\thesubsection}{\arabic{section}.\arabic{subsection}}

\renewcommand{\section}{%
	\@startsection{section}{1}{\z@}%
	{2.0ex plus 0.5ex minus 0.2ex}%
	{1.0ex plus 0.2ex}%
	{\normalfont\Large\bfseries}%
}
\renewcommand{\subsection}{%
	\@startsection{subsection}{2}{\z@}%
	{2.0ex plus 0.5ex minus 0.2ex}%
	{1.0ex plus 0.2ex}%
	{\normalfont\large\bfseries}%
}
\newcommand{\SIsection}[1]{%
	\vspace{2ex}%
	{\Large\bfseries #1\par}%
	\vspace{1ex}%
}

\newcommand{\SIsubsection}[1]{%
	\vspace{2ex}%
	{\large\bfseries #1\par}%
	\vspace{1.5ex}%
}

\makeatother
\clearpage

\vspace*{5\baselineskip}

\begin{center}
	{\large\bfseries Supplementary Material for}
	\vspace{3em}		
	
	{\Large\Large\bfseries
		Planar Nanofluidic Memristors Enabled by Surface Charge Gradient
	}
	\vspace{3em}
	
	{\normalsize
		Zhongyuan Zhao, Ziyi Yin, Chudi Qi, Yuheng Li, Shoushan Fan, Qunqing Li, and Yang Wei$^*$
	}
	\vspace{1.5em}
	
	{\small
		Department of Physics and
		Tsinghua-Foxconn Nanotechnology Research Center, State Key Laboratory of Low-Dimensional Quantum Physics, Tsinghua University, Beijing 100084, China.
	}
	\vspace{1.5em}

\end{center}

\vspace{1em}

% ================================
% Table of Contents
% ================================

\setstretch{2.5}
\begin{center}

	{\large\bfseries Contents}
	
	\vspace{1em}
	
	Nondimensionalization procedures \dotfill 1
	
	Customized finite difference method \dotfill 4
	
	Simulated results \dotfill 11
\end{center}
\thispagestyle{empty}
\clearpage

\setcounter{page}{1}
% ================================
% Supplemental Material starts here
% ================================

\setstretch{1.0}

\begin{figure*}[htbp!!!!!]
			\includegraphics[width=0.68\textwidth]{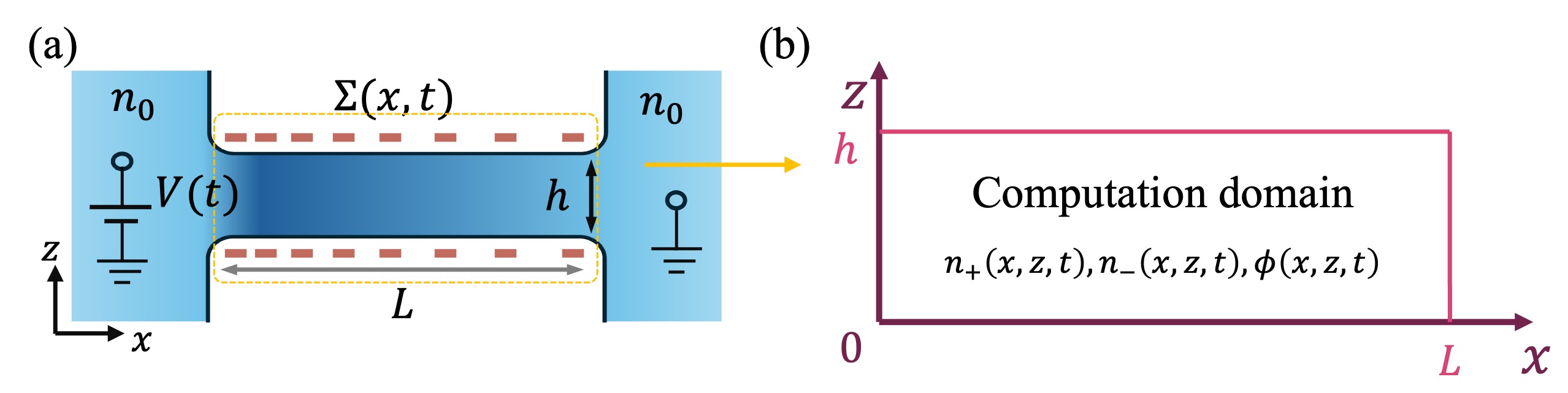}% Here is how to import EPS art
			\caption{Device configuration and computation domain.}
			\label{fig:Figure S1}
		\end{figure*}

\SIsection{1. Nondimensionalization procedures}
	\SIsubsection{1.1 Basic configuration}
		The device configuration is shown in Fig. \ref{fig:Figure S1}(a). The channel is in a nano-slit geometry, which is assumed to extend infinitely in the y-direction. The channel has a length $L$ and a height $h$, and is connected to reservoirs with ion concentration $n_0$ at both ends. In the presence of surface charge $\Sigma(x,t)$ at the channel wall, and under an applied voltage $V$, we aim to solve the evolution of ion concentration $n_\pm(x,z,t)$ and electrostatic potential $\varphi(x,z,t)$, through Poisson-Nernst-Planck equations.
		
		To formulate concise mathematical forms and efficient solving algorithms, we choose a simplistic square computation domain, as shown in Fig. \ref{fig:Figure S1}(b). With such treatment, the computation complexity can be reduced and a systematic exploration of PDE solution over the parameter space can be easily deployed. In the following sections, the limitations of this selection of computation domain will be discussed and the PDE solutions in this set will be verified against known concrete solutions.
	
	\SIsubsection{1.2 Basic Poisson-Nernst-Planck equations}
	We consider two ionic species with valences $z_+$ and $z_-$ and number densities $n_+(x,z,t)$ and $n_-(x,z,t)$. Without specifying, usually $z_+=1$ and $z_-=-1$.
	The ionic fluxes are given by the Nernst--Planck equations
	\begin{equation}
		\mathbf{J}_\alpha
		=
		- D \left(
		\nabla n_\alpha
		+ \frac{z_\alpha e}{k_\mathrm{B}T} n_\alpha \nabla \varphi
		\right),
		\qquad \alpha\in\{+,-\}.
		\label{eq:NP euqation}
	\end{equation}
	The continuity equations read
	\begin{equation}
		\frac{\partial n_\alpha}{\partial t}
		+
		\nabla\cdot\mathbf{J}_\alpha
		=
		0.
		\label{eq:continuity euqation}
	\end{equation}
	The electrostatic potential $\varphi(x,z,t)$ satisfies Poisson's equation
	\begin{equation}
		\nabla^2 \varphi
		=
		-\frac{e}{\varepsilon}
		\left(
		z_+ n_+ + z_- n_-
		\right).
		\label{eq:Poisson euqation}
	\end{equation}
	The flux expression and Poisson equation are mutually consistent and correspond
	to a standard mean-field PNP treatment.

	The Nernst--Planck flux in Eq.~\eqref{eq:NP euqation} contains two distinct contributions: 
	(i) diffusion driven by concentration gradients ($\nabla n_\alpha$), and 
	(ii) electromigration driven by the local electric field ($n_\alpha\nabla\varphi$).
	The coupled Poisson equation in Eq.~\eqref{eq:Poisson euqation} closes the system by determining the electrostatic potential self-consistently from the charge density, thereby accounting for Debye screening and electric double-layer formation near the charged interfaces.
	
	The present formulation adopts a mean-field continuum description of the electrolyte.
	In particular, ion--ion correlations, finite ion size (steric effects), and short-range hydration interactions are neglected. These approximations are generally valid for dilute electrolytes and moderate surface potentials, and they provide a minimal yet physically rich framework for exploring the interplay between confinement, screening, and surface-charge-induced modulation.
	
	\SIsubsection{1.3 Introduction of characteristic length and time}
	We introduce the Debye length
	\begin{equation}
		l_\mathrm{D} = \sqrt{\frac{\varepsilon k_\mathrm{B} T}{2 e^2 n_0}},
	\end{equation}
	and the diffusive time scale
	\begin{equation}
		\tau_\mathrm{D} = \frac{l_\mathrm{D}^2}{D}=\frac{\varepsilon k_\mathrm{B}T}{2n_0e^2D}=\frac{\varepsilon}{2n_0e\mu }=\frac{\varepsilon}{\sigma_0}.
	\end{equation}
	Dimensionless variables are defined as
	\begin{equation}
		\bar{x} = \frac{x}{l_\mathrm{D}},
		\qquad
		\bar{z} = \frac{z}{l_\mathrm{D}},
		\qquad
		\bar{t} = \frac{t}{\tau_\mathrm{D}},
		\qquad
		\Phi = \frac{e\varphi}{k_\mathrm{B} T},
		\qquad
		\bar{n}_\alpha = \frac{n_\alpha}{n_0}.
	\end{equation}
	The dimensionless gradient and time derivative operators are
	\begin{equation}
		\bar{\nabla} = l_\mathrm{D} \nabla,
		\qquad
		\partial_{\bar{t}} = \tau_\mathrm{D}\partial_t.
	\end{equation}
	
	The Debye length $l_\mathrm{D}$ sets the thickness of the diffuse layer near charged boundaries. 
	The time scale $\tau_\mathrm{D}=l_\mathrm{D}^2/D$ is the characteristic diffusion time across one Debye length and therefore controls the relaxation of ionic charge density near interfaces.
	In addition, the geometry enters through the two dimensionless ratios $\theta=L/h$ and $\gamma=h/4l_\mathrm{D}$, which quantify the channel aspect ratio and the relative strength of confinement compared with screening.
	The common value for $D$ is about $2\times 10^{-9} \mathrm{m}^2/\mathrm{s}$ (for $\mathrm{K}^+$ and $\mathrm{Cl}^-$ in dilute aqueous  solution), and the Debye length is usually of nanometer scale. Hence $\tau_\mathrm{D}$ usually has a ns scale. 

	\SIsubsection{1.4 Dimensionless equations and boundary conditions}
		In dimensionless form, the PNP system becomes
		\begin{align}
			\partial_{\bar{t}}\bar{n}_+
			&=-\bar{\nabla}\cdot\bar{\mathbf{J}}_+=\bar{\nabla}\cdot\left(\bar{\nabla}\bar{n}_{+}+z_{+} \bar{n}_{+} \bar{\nabla}\Phi\right),
			\\
			\partial_{\bar{t}}\bar{n}_-
			&=-\bar{\nabla}\cdot\bar{\mathbf{J}}_-=\bar{\nabla}\cdot\left(\bar{\nabla}\bar{n}_{-}+z_{-} \bar{n}_{-} \bar{\nabla}\Phi\right),
			\\
			\bar{\nabla}^2\Phi
			&=
			-\frac{1}{2}\left(z_+ \bar{n}_+ + z_- \bar{n}_-\right).
		\end{align}
		The computational domain is
		\[
		(\bar{x},\bar{z})\in[0,4\theta\gamma]\times[0,4\gamma],
		\qquad
		\theta = \frac{L}{h},
		\quad
		\gamma = \frac{h}{4l_\mathrm{D}}.
		\]
		The electrostatic potential satisfies
		\begin{equation}
			\Phi(\bar{x}=0,\bar{z},\bar{t}) = \frac{eV(\bar{t})}{k_\mathrm{B}T},
			\qquad
			\Phi(\bar{x}=\theta\gamma,\bar{z},\bar{t}) = 0.
		\end{equation}
		The normal electric field at the channel-electrolyte interface generates from the surface charge $\Sigma(\bar{x})$, and it is imposed in a dimensionless way:
		\begin{equation}
			-\partial_{\bar{z}}\Phi(x,\bar{z}=0,\bar{t})
			=
			\partial_{\bar{z}}\Phi(x,\bar{z}=\gamma,\bar{t})
			=
			2\chi(x,\bar{t}),
		\end{equation}
		where
		\begin{equation}
			\chi(\bar{x},\bar{t})
			=
			\frac{l_\mathrm{D}}{l_\mathrm{GC}(\bar{x},\bar{t})},
			\qquad
			l_\mathrm{GC}(\bar{x},\bar{t})
			=
			\frac{\varepsilon k_\mathrm{B} T}{2 e^2 \Sigma(\bar{x},\bar{t})}.
		\end{equation}
		
		To close the Poisson--Nernst--Planck (PNP) system, boundary conditions must be specified not only for the
		electrostatic potential $\Phi$, but also for the ionic densities $\bar{n}_\alpha$ (equivalently, for the
		ionic fluxes). In the present device geometry, the channel is connected to macroscopic ionic reservoirs at
		its left and right ends ($x=0$ and $x=L$). We therefore adopt the simplest ``strong reservoir'' approximation,
		namely Dirichlet concentration boundary conditions:
		\begin{equation}
			\bar{n}_\alpha(\bar{x}=0,\bar{z},\bar{t}) = 1,
			\qquad
			\bar{n}_\alpha(\bar{x}=4\theta\gamma,\bar{z},\bar{t}) = 1,
			\qquad
			\alpha\in\{+,-\},
			\label{eq:bc_reservoir_dirichlet}
		\end{equation}
		which correspond to fixed bulk concentrations $n_\alpha=n_0$ maintained at the channel openings.
		Physically, this assumes that the external reservoirs are sufficiently large and well-mixed so that their
		concentrations do not change under the ionic currents induced by the applied driving $V(\bar{t})$. In this framework, the reservoirs act as ideal particle baths with infinite supply and	perfect relaxation.
		
		At the upper and lower channel boundaries ($z=0$ and $z=h$), we neglect Faradaic charge-transfer reactions and assume that the dielectric walls are ion-impermeable. The appropriate condition is therefore a
		\emph{blocking (no-normal-flux) boundary condition} for each ionic species:
		\begin{equation}
			\mathbf{J}_\alpha\cdot\hat{\mathbf{n}} = 0
			\qquad \text{on } \bar{z}=0,\;\gamma,
			\qquad
			\alpha\in\{+,-\},
			\label{eq:bc_blocking_flux_dimensional}
		\end{equation}
		where $\hat{\mathbf{n}}$ is the outward unit normal and the dimensionless Nernst--Planck flux is
		\begin{equation}
			\bar{\mathbf{J}}_\alpha
			=
			-\left(\bar{\nabla}\bar{n}_\alpha + z_\alpha \bar{n}_\alpha \bar{\nabla}\Phi\right).
			\label{eq:flux_dimensionless_again}
		\end{equation}
		Since $\hat{\mathbf{n}}=\mp\hat{\mathbf{z}}$ on the two walls, Eq.~\eqref{eq:bc_blocking_flux_dimensional}
		is equivalently expressed as
		\begin{equation}
			\left.
			\partial_{\bar{z}}\bar{n}_\alpha + z_\alpha \bar{n}_\alpha\,\partial_{\bar{z}}\Phi
			\right|_{\bar{z}=0}=0,
			\qquad
			\left.
			\partial_{\bar{z}}\bar{n}_\alpha + z_\alpha \bar{n}_\alpha\,\partial_{\bar{z}}\Phi
			\right|_{\bar{z}=\gamma}=0.
			\label{eq:bc_blocking_flux_explicit}
		\end{equation}
		These conditions enforce that ions cannot cross the channel--dielectric interfaces; however, tangential transport along z direction remains allowed, and the surface-charge-induced normal electric field ($\partial_{\bar{z}}\Phi \neq 0$) can still strongly modulate the near-wall ion distributions through
		electromigration.
		
		Despite its simplicity and robustness, this boundary-condition set represents an idealization with several
		limitations:
		(i) imposing fixed concentrations at $\bar{x}=0$ and $\bar{x}=4\theta\gamma$ suppresses any concentration
		evolution \emph{inside the reservoirs} and therefore cannot describe polarization layers extending into the
		external baths; (ii) the Dirichlet reservoir assumption corresponds to infinitely fast mixing, which may
		overestimate the ion supply under strong driving or near limiting-current regimes; (iii) the blocking-wall
		condition excludes Faradaic reactions and surface conduction through chemically active boundaries, and thus
		cannot capture current carried by charge transfer or reaction-limited kinetics at the interfaces.
		If such effects are important, more refined open-boundary models (e.g.\ Robin/mixed conditions) or explicit
		reservoir domains, as well as reactive boundary conditions, should be incorporated.
	\clearpage
\SIsection{2. Customized finite difference method}
	\SIsubsection{2.1 Reasons for customizing solving methods}
	The decision to develop a customized PDE solver, rather than relying on established commercial or open-source packages, is grounded in a combination of scientific, computational, and practical necessities. 
	
	Firstly, custom development allows for problem-specific specialization. General-purpose software is inherently designed to cover a broad spectrum of physics and numerical methods, which inevitably introduces computational overhead and compromises on performance for any single application. By tailoring the solver exclusively to our governing equations and boundary conditions, we can embed optimized linear solvers, and problem-aware discretization schemes. This specialization means we can aggressively prune redundant generalities and dynamically adjust the algorithm based on real-time feedback from practical simulations—for example, by adaptively tuning the error tolerances, thereby drastically enhancing computational throughput compared to a ``one-size-fits-all" black box.
	
	Secondly, autonomous algorithm development fundamentally facilitates high-throughput parametric studies. In typical research scenarios, we require sweeping across high-dimensional parameter spaces (e.g., varying initial conditions, material coefficients, or device configurations) to map out regime diagrams. While off-the-shelf software often provides powerful single-run capabilities, they are sometimes difficult to embed into large-scale automated workflows. 
	
	Thirdly, custom development guarantees complete transparency and comprehensive first-hand grasp of the underlying physics. When using a commercial solver, the numerical implementation of the physical terms—including how boundary conditions are weakly enforced, how nonlinearities are linearized, and how source terms are integrated—is often obscured behind compiled code or poorly documented kernels. This opacity becomes a critical liability when simulations produce non-physical artifacts or convergence failures. By writing the solver on our own, we retain absolute control over the solving process. Consequently, we derive far deeper insights into the model's fidelity.

\begin{figure}[htbp]
			\centering
			\includegraphics[width=0.8\linewidth]{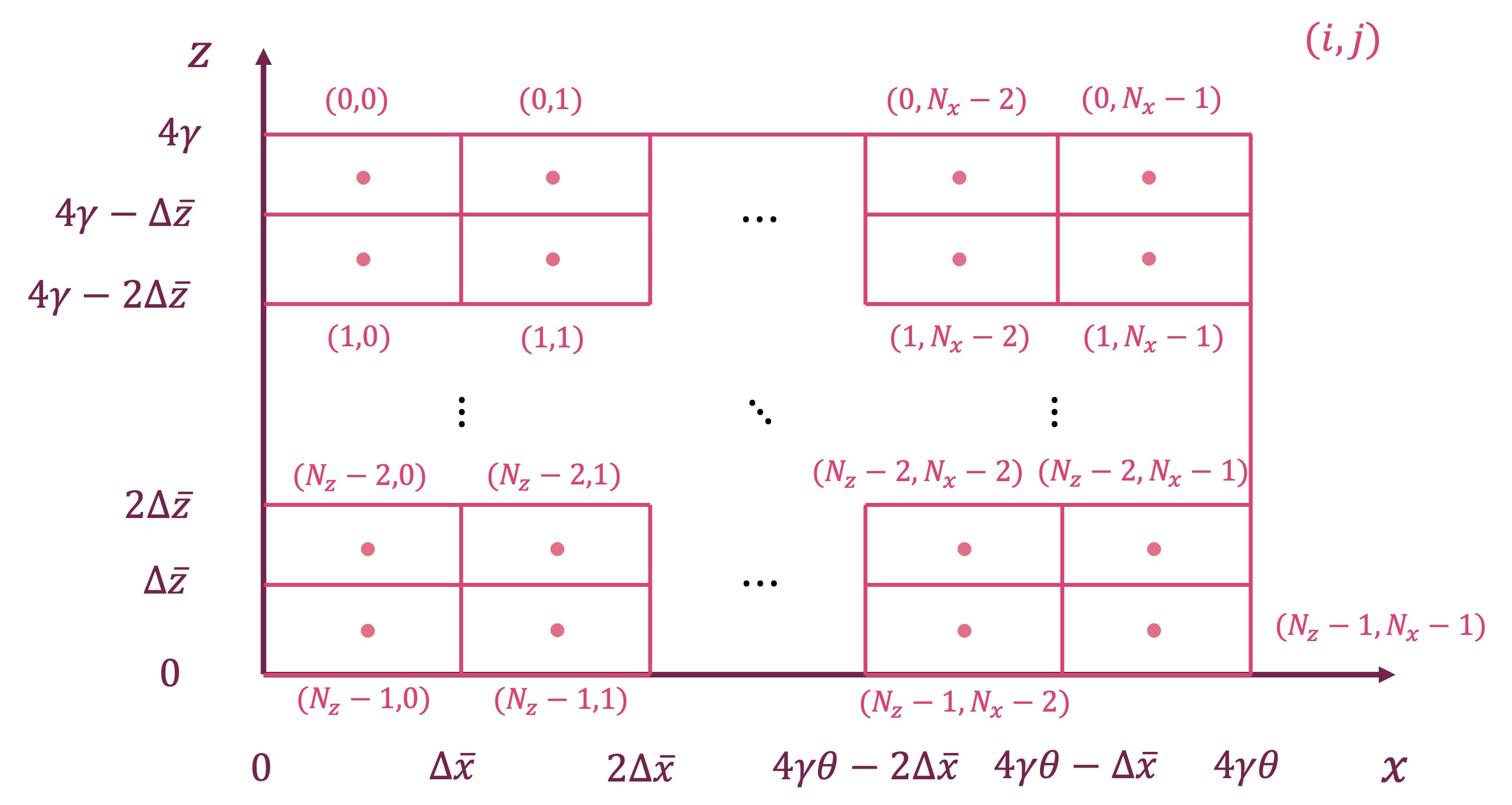}
			\caption{The uniform grid of spatial computational domain.}
			\label{fig:Discrete_spatial_grid}
		\end{figure}

	\SIsubsection{2.2 Uniform discretized grid}
		The partial differential equations to solve include the Poisson equation $\bar{\nabla}^2\Phi = -\frac{1}{2}(z_+\bar{n}_+ + z_-\bar{n}_-)$ and the Smoluchowski equation $	\partial_{\bar{t}}\bar{n}_\alpha=\bar{\nabla}\cdot\Bigl(\bar{\nabla}\bar{n}_\alpha +z_\alpha\bar{n}_+\bar{\nabla}\Phi\Bigr)$. And the latter equation can be rewirtten by making using of the former:
		\begin{equation}
			\partial_{\bar{t}}\bar{n}_\alpha=\bar{\nabla}^2\bar{n}_\alpha+z_\alpha\Bigl(\partial_{\bar{x}}\bar{n}_\alpha\cdot\partial_{\bar{x}}\Phi+\partial_{\bar{z}}\bar{n}_\alpha\cdot\partial_{\bar{z}}\Phi\Bigr)-\frac{1}{2}z_\alpha\bar{n}_\alpha\Bigl(z_+\bar{n}_++z_-\bar{n}_-\Bigr).
		\end{equation}
		The spatial and temporal domains are uniformly discretized as below:
		\begin{eqnarray}
			&i=0, 1, ..., N_z-1;\quad j=0, 1, ..., N_x-1;\quad k=0, 1, ..., N_t -1\\
				&N_z\cdot\Delta\bar{z}=4\theta\gamma;\quad 	N_x\cdot\Delta\bar{x}=4\gamma;\quad N_t\cdot\Delta\bar{t}=\bar{T}. 
		\end{eqnarray}
		The spatial and temporal coordinates are now labelled by $i, j, k$:
		\begin{equation}
			(\bar{x},\bar{z})_{i,j}=\Bigl(\frac{1}{2}\Delta\bar{x}+j\cdot\Delta\bar{x},(N_z-\frac 1{2})\Delta\bar{z}-i\cdot\Delta\bar{z}\Bigr); \quad \bar{t}_k=k\cdot\Delta\bar{t}
		\end{equation}
		The spatial discrete grid are illustrated in Fig.~\ref{fig:Discrete_spatial_grid}.
		
		Hence what are solved from the PDEs are the discrete values of $\Phi$ and $\bar{n}_\alpha$ on each grid sites:
		\begin{equation}
			\Phi(\bar{x},\bar{z},\bar{t})\rightarrow\Phi_{i,j,k},\quad 	\bar{n}_\alpha(\bar{x},\bar{z},\bar{t})\rightarrow(\bar{n}_\alpha)_{i,j,k}
		\end{equation}
	\SIsubsection{2.3 Discretization of operators}
		The PDEs to be solved contain differential operators like $\bar{\nabla}^2$, $\partial_{\bar{x}}$ and $\partial_{\bar{z}}$. Since discretizations have been introduced to the spatial computational domain $[0,4\gamma]\times[0,4\gamma\theta]$, discretized versions of the differential operators should also be found. Here partial derivatives are converted to linear matrix-vector-multiplications (MVM). The values of $\Phi$ and $\bar{n}_\alpha$ on discrete grid sites are aranged to vectors and the differential operators $\bar{\nabla}^2$, $\partial_{\bar{x}}$ and $\partial_{\bar{z}}$ corresponds to matrices $A$, ${}^x\!B$ and  ${}^z\!B$. Since both $\Phi$ and $\bar{n}_\alpha$ adopt mixed types of boundary conditions (Dirichlet type at left or right boundaries, while Riemann type at the upper or lower boundaries:
		\begin{eqnarray}
			\Phi(\bar{x}=0,\bar{z},\bar{t})=\Phi_{\mathrm{ds}}(\bar{t}),&\quad&\Phi(\bar{x}=4\theta\gamma,\bar{z},\bar{t})=0,\\
			\partial_{\bar{z}}\Phi(\bar{x},\bar{z}=0,\bar{t})=-2\chi(\bar{t}),&\quad&\partial_{\bar{z}}\Phi(\bar{x},\bar{z}=4\gamma,\bar{t})=2\chi(\bar{t}),\\ 			
			\bar{n}_\alpha(\bar{x}=0,\bar{z},\bar{t})=\bar{n}_{\alpha,\mathrm{L}},&\quad&\bar{n}_\alpha(\bar{x}=4\theta\gamma,\bar{z},\bar{t})=\bar{n}_{\alpha,\mathrm{R}},\\
			\partial_{\bar{z}}\bar{n}|_{\bar{z}=0}=2z_\alpha\bar{n}_\alpha(\bar{x},\bar{z}=0,\bar{t})\chi(\bar{t}),&\quad 			&\partial_{\bar{z}}\bar{n}|_{\bar{z}=4\gamma}=-2z_\alpha\bar{n}_\alpha(\bar{x},\bar{z}=4\gamma,\bar{t})\chi(\bar{t}),
		\end{eqnarray}
		a unified problem can be formulated to be finding the linearized forms of $\bar{\nabla}^2\phi$, $\partial_{\bar{x}}\phi$ and $\partial_{\bar{z}}\phi$ when provided with $\phi(\bar{x},\bar{z})$ and its boundary conditions $\phi_\mathrm{L}$, $\phi_\mathrm{R}$, $\phi_\mathrm{U}^\prime$ and $\phi_\mathrm{D}^\prime$:
		\begin{eqnarray}
			\label{eq:MVM_A_b_definition}
			\left(\bar{\nabla}^2\phi\right)_{i,j}&=&\sum_{i^\prime\!,j^\prime}A^{i^\prime\!,j^\prime}_{i,j}(\mathrm{Dim}\phi)\cdot\phi_{i^\prime\!,j^\prime}+b_{i,j}(\mathrm{Dim}\phi,\mathrm{Boun}\phi),\\
			\label{eq:MVM_Bx_cx_definition}
			\left(\partial_{\bar{x}}\phi\right)_{i,j}&=&\sum_{i^\prime\!,j^\prime}{}^x\!B^{i^\prime\!,j^\prime}_{i,j}(\mathrm{Dim}\phi)\cdot\phi_{i^\prime\!,j^\prime}+{}^x\!c_{i,j}(\mathrm{Dim}\phi,\mathrm{Boun}\phi),\\		
			\label{eq:MVM_Bz_cz_definition}
			\left(\partial_{\bar{z}}\phi\right)_{i,j}&=&\sum_{i^\prime\!,j^\prime}{}^z\!B^{i^\prime\!,j^\prime}_{i,j}(\mathrm{Dim}\phi)\cdot\phi_{i^\prime\!,j^\prime}+{}^z\!c_{i,j}(\mathrm{Dim}\phi,\mathrm{Boun}\phi).
		\end{eqnarray}
		$\mathrm{Dim}\phi$ represents the dimension information of $\phi$, including $N_x$, $N_z$, $\Delta\bar{x}$ and $\Delta\bar{z}$. $\mathrm{Boun}\phi$ represents the boundary condition information of $\phi$, including $\phi_\mathrm{L}$, $\phi_\mathrm{R}$, $\phi_\mathrm{U}^\prime$ and $\phi_\mathrm{D}^\prime$. Special attentions should be paid to the edges and corners of the spatial grid, so that the boundary conditions can be correctly incorporated to the vectors $b_{i,j}(\mathrm{Dim}\phi,\mathrm{Boun}\phi)$, ${}^x\!c_{i,j}(\mathrm{Dim}\phi,\mathrm{Boun}\phi)$ and ${}^z\!c_{i,j}(\mathrm{Dim}\phi,\mathrm{Boun}\phi)$. The $(\bar{x},\bar{z})$ coordinates and the $(i,j)$ labels for the edges (L for left, R for right, U for upper, and L for lower) and corners (LU for upper left, and so on) are listed in Table.~\ref{table1:positions_of_boundaries_corners}.
		
		\begin{table}[h]
			\centering
			\renewcommand{\arraystretch}{1.5}
			\setlength{\tabcolsep}{8.9pt}
			\begin{tabular}{|c|c|c|c|c|}
				\hline
				position & $\bar{x}$ & $\bar{z}$ & $i$ & $j$ \\
				\hline
				L        & $0$  & $0\sim4\gamma$  & $1, 2, .. , N_z-2$ & $0$  \\
				\hline
				R        & $4\gamma\theta$  & $0\sim4\gamma$  & $1, 2, .. , N_z-2$  & $N_x-1$  \\
				\hline
				U        & $0\sim4\gamma\theta $ & $4\gamma$  & $0$ & $1, 2, ..., Nx-2$  \\
				\hline
				D        & $0\sim4\gamma\theta $  & $0$  & $N_z-1$  &  $1, 2, ..., Nx-2 $\\
				\hline
				LU       &  $0$ & $4\gamma$  & $0$  & $0$  \\
				\hline
				LD       & $0$  &  $0$ &  $N_z-1$ & $0$  \\
				\hline
				RU       & $4\gamma\theta$  & $4\gamma$  & $0$  & $N_x-1$  \\
				\hline
				RD       & $4\gamma\theta$ & $0$ & $N_z-1$ & $N_x-1$ \\
				\hline
				Inner&$0\sim4\gamma\theta$&$0\sim4\gamma$&$1, 2, .. , N_z-2$&$1, 2, .. , N_x-2$\\
				\hline
			\end{tabular}
			\caption{Coordinates and discrete labels for the boundaries and corners}
			\label{table1:positions_of_boundaries_corners}
		\end{table}
		\begin{figure}[htbp]
			\centering
			\includegraphics[width=0.7\linewidth]{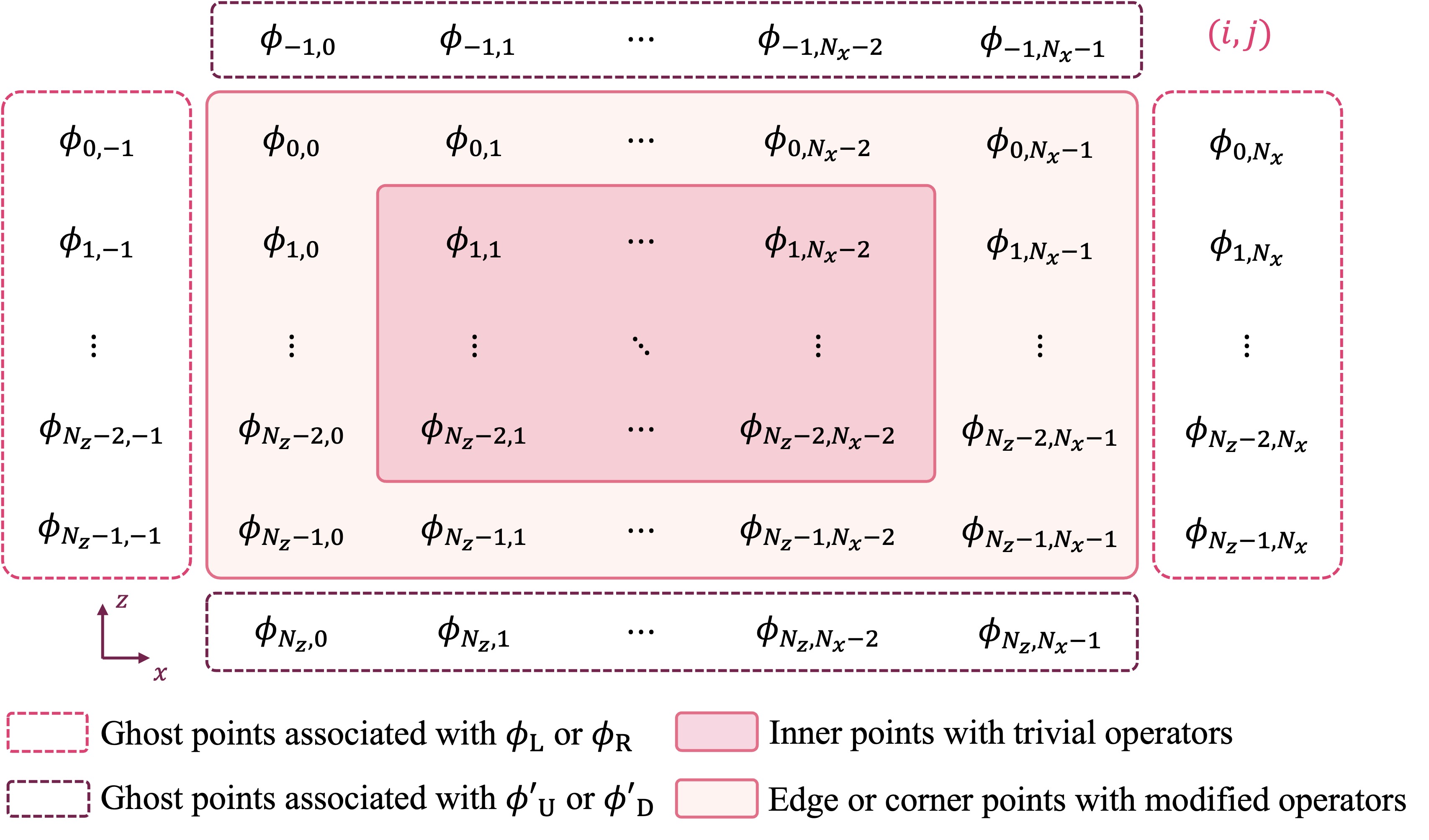}
			\caption{Introduction of ghost points and their associations to boundary conditions}
			\label{fig:Ghost_point_scheme}
		\end{figure}
		Additionally, ghost points are introduced to associate with $\phi_\mathrm{L}$, $\phi_\mathrm{R}$, $\phi_\mathrm{U}^\prime$ and $\phi_\mathrm{D}^\prime$. This scheme is illustrated in Fig.~\ref{fig:Ghost_point_scheme}. Below the discretized and linearized forms for $\bar{\nabla}^2\phi$, $\partial_{\bar{x}}\phi$ and $\partial_{\bar{z}}\phi$ are derived with second-order accuracy:
		\begin{eqnarray}
			\label{eq:Discrete_Laplacian_operator}
			\left(\bar{\nabla}^2\phi\right)_{i,j}&=&\frac{\phi_{i,j-1}-2\phi_{i,j}+\phi_{i,j+1}}{\Delta\bar{x}^2}+\frac{\phi_{i-1,j}-2\phi_{i,j}+\phi_{i+1,j}}{\Delta\bar{z}^2},\\
			\left(\partial_{\bar{x}}\phi\right)_{i,j}&=&\frac{\phi_{i,j+1}-\phi_{i,j-1}}{2\Delta\bar{x}},\\			\left(\partial_{\bar{z}}\phi\right)_{i,j}&=&\frac{\phi_{i-1,j}-\phi_{i+1,j}}{2\Delta\bar{z}}.
		\end{eqnarray}
		And the boundary conditions are applied in the forms of:
		\begin{eqnarray}
			\label{eq:LR_boundary_conditions}
			\phi_{i,-1}=\phi_{\mathrm{L}},\qquad\phi_{i,N_x}=\phi_{\mathrm{R}},\qquad&i=0, 1, ..., N_z-1\\
			\label{eq:UD_boundary_conditions}
			\frac{\phi_{-1,j}-\phi_{1,j}}{2\Delta\bar{z}}=\phi_{\mathrm{U}}^\prime,\qquad\frac{\phi_{N_z-2,j}-\phi_{N_z,j}}{2\Delta\bar{z}}=\phi_{\mathrm{D}}^\prime,\qquad&j=0, 1, ..., N_x-1
		\end{eqnarray} 		
		Hence according to eqs.~(\ref{eq:MVM_A_b_definition})-(\ref{eq:MVM_Bz_cz_definition}), the MVM forms are:
		\begin{eqnarray}
			\left(\bar{\nabla}^2\phi\right)_{i,j}&=&A_{i,j}^{i,j}\phi_{i,j}+A_{i,j}^{i,j-1}\phi_{i,j-1}+A_{i,j}^{i,j+1}\phi_{i,j+1}+A_{i,j}^{i-1,j}\phi_{i-1,j}+A_{i,j}^{i+1,j}\phi_{i+1,j}+b_{i,j},\\
			\left(\partial_{\bar{x}}\phi\right)_{i,j}&=&{}^x\!B^{i,j-1}_{i,j}\phi_{i,j-1}+{}^x\!B^{i,j+1}_{i,j}\phi_{i,j+1}+{}^x\!c_{i,j},\\			\left(\partial_{\bar{z}}\phi\right)_{i,j}&=&{}^z\!B^{i-1,j}_{i,j}\phi_{i-1,j}+{}^z\!B^{i+1,j}_{i+1,j}\phi_{i,j+1}+{}^z\!c_{i,j}.
		\end{eqnarray}
		Now  \emph{let $\alpha=1/\Delta\bar{x}$ and $\beta=1/\Delta\bar{z}$ } and find the specific values of matrix and vector elements. Take $\left(\bar{\nabla}^2\phi\right)_{i,j}$ as an example, and for inner grid sites ($i=1, 2, ..., N_z-2,\quad j=1, 2, ..., N_x-2$), from eq.~(\ref{eq:Discrete_Laplacian_operator}):
		\begin{equation}
			\left(\bar{\nabla}^2\phi\right)_{i,j}=-2(\alpha^2+\beta^2)\phi_{i,j}+\alpha^2\phi_{i,j-1}+\alpha^2\phi_{i,j+1}+\beta^2\phi_{i-1,j}+\beta^2\phi_{i+1,j}.
		\end{equation}
		Hence the matrix and vector elements are found to be:
		\begin{equation}
			A_{i,j}^{i,j}=-2(\alpha^2+\beta^2),\quad A_{i,j}^{i,j-1}=\alpha^2,\quad A_{i,j}^{i,j+1}=\alpha^2,\quad A_{i,j}^{i-1,j}=\beta^2,\quad A_{i,j}^{i+1,j}=\beta^2,\quad b_{i,j}=0
		\end{equation}
		For sites on the grid edges, take the left edge ($\mathrm{L}: j=0, i=1, 2, ..., N_z-2$) as an example, and from eqs.~(\ref{eq:Discrete_Laplacian_operator}) and (\ref{eq:LR_boundary_conditions}):
		\begin{equation}
			\left(\bar{\nabla}^2\phi\right)_{i,j}=-2(\alpha^2+\beta^2)\phi_{i,j}+\alpha^2\phi_{\mathrm{L}}+\alpha^2\phi_{i,j+1}+\beta^2\phi_{i-1,j}+\beta^2\phi_{i+1,j}.
		\end{equation}
		Hence the matrix and vector elements are found to be:
		\begin{equation}
			A_{i,j}^{i,j}=-2(\alpha^2+\beta^2),\quad A_{i,j}^{i,j+1}=\alpha^2,\quad A_{i,j}^{i-1,j}=\beta^2,\quad A_{i,j}^{i+1,j}=\beta^2,\quad b_{i,j}=\alpha^2\phi_{\mathrm{L}},
		\end{equation}
		and $A_{i,j}^{i,j-1}$ does not exist. For sites on the grid corners, take the upper left corner ($\mathrm{L}: i=0, j=0$) as an example, and from eqs.~(\ref{eq:Discrete_Laplacian_operator}), (\ref{eq:LR_boundary_conditions}) and (\ref{eq:UD_boundary_conditions}):
		\begin{equation}
			\left(\bar{\nabla}^2\phi\right)_{i,j}=-2(\alpha^2+\beta^2)\phi_{i,j}+\alpha^2\phi_{\mathrm{L}}+\alpha^2\phi_{i,j+1}+\beta^2(\frac{2}{\beta}\phi_{\mathrm{U}}^\prime+\phi_{i+1,j})+\beta^2\phi_{i+1,j}.
		\end{equation}
		Hence the matrix and vector elements are found to be:
		\begin{equation}
			A_{i,j}^{i,j}=-2(\alpha^2+\beta^2),\quad A_{i,j}^{i,j+1}=\alpha^2,\quad A_{i,j}^{i+1,j}=2\beta^2,\quad b_{i,j}=\alpha^2\phi_{\mathrm{L}}+2\beta\phi_{\mathrm{U}}^\prime,
		\end{equation}
		and $A_{i,j}^{i,j-1}$, $A_{i,j}^{i-1,j}$ do not exist. Likewise, all the matrix elements for $A$, ${}^x\!B$ and ${}^z\!B$, and all the vector elements for $b$, ${}^x\!c$ and ${}^z\!c$ can be found, as listed in Tables.~(\ref{table:elements_A_b}) and (\ref{table:elements_B_c}).
		\begin{table}[]
			\centering
			\renewcommand{\arraystretch}{1.5}
			\setlength{\tabcolsep}{8.9pt}
			\begin{tabular}{|c|c|c|c|c|c|c|}
				\hline
				& $A^{i,j}_{i,j}$ & $A^{i,j-1}_{i,j}$ & $A^{i,j+1}_{i,j}$ & $A^{i-1,j}_{i,j}$&$A^{i+1,j}_{i,j}$ &$b_{i,j}$ \\ \hline
				Inner & $-2(\alpha^2+\beta^2)$&$\alpha^2$&$\alpha^2$&$\beta^2$&$\beta^2$&0\\ \hline
				L        &  $-2(\alpha^2+\beta^2)$ & /  &  $\alpha^2$ &$\beta^2$&$\beta^2$&$\alpha^2\phi_{\mathrm{L}}$   \\ \hline
				R        & $-2(\alpha^2+\beta^2)$  & $\alpha^2$  & /  &$\beta^2$&$\beta^2$&$\alpha^2\phi_{\mathrm{R}}$   \\ \hline
				U        &  $-2(\alpha^2+\beta^2)$ & $\alpha^2$  &  $\alpha^2$ &/&$2\beta^2$&$2\beta\phi_{\mathrm{U}}^\prime$   \\ \hline
				D        & $-2(\alpha^2+\beta^2)$  & $\alpha^2$  & $\alpha^2$  &$2\beta^2$&/& $-2\beta\phi_{\mathrm{D}}^\prime$  \\ \hline
				LU       &  $-2(\alpha^2+\beta^2)$ & /  &  $\alpha^2$ &/&$2\beta^2$&$\alpha^2\phi_{\mathrm{L}}+2\beta\phi_{\mathrm{U}}^\prime$    \\ \hline
				LD       &  $-2(\alpha^2+\beta^2)$ &  / & $\alpha^2$  & $2\beta^2$&/ &$\alpha^2\phi_{\mathrm{L}}-2\beta\phi_{\mathrm{D}}^\prime$  \\ \hline
				RU       &  $-2(\alpha^2+\beta^2)$ & $\alpha^2$  & /  &/&$2\beta^2$&$\alpha^2\phi_{\mathrm{R}}+2\beta\phi_{\mathrm{U}}^\prime$    \\ \hline
				RD       &$-2(\alpha^2+\beta^2)$   &  $\alpha^2$ &  / &$2\beta^2$&/ &$\alpha^2\phi_{\mathrm{R}}-2\beta\phi_{\mathrm{D}}^\prime$   \\ \hline
			\end{tabular}
			\caption{Elements for matrix $A$ and vector $b$.}
			\label{table:elements_A_b}
		\end{table}
		\begin{table}[htbp]
			\centering
			\renewcommand{\arraystretch}{1.5}
			\setlength{\tabcolsep}{8.9pt}
			\begin{tabular}{|c|c|c|c|c|c|c|}
				\hline
				& ${}^x\!B^{i,j-1}_{i,j}$ & ${}^x\!B^{i,j+1}_{i,j}$ & ${}^z\!B^{i-1,j}_{i,j}$ & ${}^z\!B^{i+1,j}_{i,j}$&${}^x\!c_{i,j}$ &${}^z\!c_{i,j}$ \\ \hline
				Inner & $-\frac{1}{2}\alpha$&$\frac{1}{2}\alpha$&$\frac{1}{2}\beta$ &$-\frac{1}{2}\beta$&$0$&$0$\\ \hline
				L        &  / & $\frac{1}{2}\alpha$ &  $\frac{1}{2}\beta$ &$-\frac{1}{2}\beta$&$-\frac{1}{2}\alpha\phi_{\mathrm{L}}$&$0$   \\ \hline
				R        & $-\frac{1}{2}\alpha$&/& $\frac{1}{2}\beta$  &$-\frac{1}{2}\beta$&$\frac{1}{2}\alpha\phi_{\mathrm{R}}$&0   \\ \hline
				U        &  $-\frac{1}{2}\alpha$ & $\frac{1}{2}\alpha$  &  / &0&0&$\phi_{\mathrm{U}}^\prime$   \\ \hline
				D        & $-\frac{1}{2}\alpha$  & $\frac{1}{2}\alpha$  & 0  &/&0& $\phi_{\mathrm{D}}^\prime$  \\ \hline
				LU       &  /& $\frac{1}{2}\alpha$  &  / &0&$-\frac{1}{2}\alpha\phi_{\mathrm{L}}$&$\phi_{\mathrm{U}}^\prime$    \\ \hline
				LD       & /& $\frac{1}{2}\alpha$ & 0  & /&$-\frac{1}{2}\alpha\phi_{\mathrm{L}}$ &$\phi_{\mathrm{D}}^\prime$  \\ \hline
				RU       &$-\frac{1}{2}\alpha$& /  & /  &0&$\frac{1}{2}\alpha\phi_{\mathrm{R}}$&$\phi_{\mathrm{U}}^\prime$  \\ \hline
				RD       &$-\frac{1}{2}\alpha$ &  /& 0 &/&$\frac{1}{2}\alpha\phi_{\mathrm{R}}$ &$\phi_{\mathrm{D}}^\prime$   \\ \hline
			\end{tabular}
			\caption{Elements for matrices ${}^x\!B$ and ${}^z\!B$, and for vectors ${}^x\!c$ and ${}^z\!c$.}
			\label{table:elements_B_c}
		\end{table}
		\begin{table}[]
			\centering
			\renewcommand{\arraystretch}{1.5}
			\setlength{\tabcolsep}{2pt}
			\begin{tabular}{|c|c|c|c|c|c|c|c|c|c|c|c|c|c|c|c|c|c|}
				\hline
				& ${}^0\!I^{i,j}_{i,j}$ & ${}^1\!I^{i,j-1}_{i,j}$ & ${}^2\!I^{i,j+1}_{i,j}$ &${}^3\!I^{i-1,j}_{i,j}$ &${}^4\!I^{i+1,j}_{i,j}$ &${}^5\!I^{i,j-1}_{i,j}$&${}^6\!I^{i,j+1}_{i,j}$&${}^7\!I^{i-1,j}_{i,j}$&${}^8\!I^{i+1,j}_{i,j}$&${}^0\!a$&${}^1\!a$&${}^2\!a$&${}^3\!a$&${}^4\!a$&${}^5\!a$&${}^6\!a$&${}^7\!a$ \\ \hline
				Inner & 1 & 1 & 1 & 1 & 1 & 1 & 1 & 1 & 1 & 0 & 0 & 0 & 0 & 0 & 0 & 0 & 0 \\ \hline
				L     & 1 & / & 1 & 1 & 1 & / & 1 & 1 & 1 & 1 & 0 & 0 & 0 & 1 & 0 & 0 & 0 \\ \hline
				R     & 1 & 1 & / & 1 & 1 & 1 & / & 1 & 1 & 0 & 1 & 0 & 0 & 0 & 1 & 0 & 0 \\ \hline
				U     & 1 & 1 & 1 & / & 2 & 1 & 1 & / & 0 & 0 & 0 & 1 & 0 & 0 & 0 & 1 & 0 \\ \hline
				D     & 1 & 1 & 1 & 2 & / & 1 & 1 & 0 & / & 0 & 0 & 0 & 1 & 0 & 0 & 0 & 1 \\ \hline
				LU    & 1 & / & 1 & / & 2 & / & 1 & / & 0 & 1 & 0 & 1 & 0 & 1 & 0 & 1 & 0 \\ \hline
				LD    & 1 & / & 1 & 2 & / & / & 1 & 0 & / & 1 & 0 & 0 & 1 & 1 & 0 & 0 & 1 \\ \hline
				RU    & 1 & 1 & / & / & 2 & 1 & / & / & 0 & 0 & 1 & 1 & 0 & 0 & 1 & 1 & 0 \\ \hline
				RD    & 1 & 1 & / & 2 & / & 1 & / & 0 & / & 0 & 1 & 0 & 1 & 0 & 1 & 0 & 1 \\ \hline
			\end{tabular}
			\caption{Predefined purely numbered matrices $I$ and vectors $a$ (the subscripts $i,j$ are neglected for $a$).}
			\label{table:elements_I_a}
		\end{table}
		
\begin{figure}[htbp]
			\centering
			\includegraphics[width=0.8\linewidth]{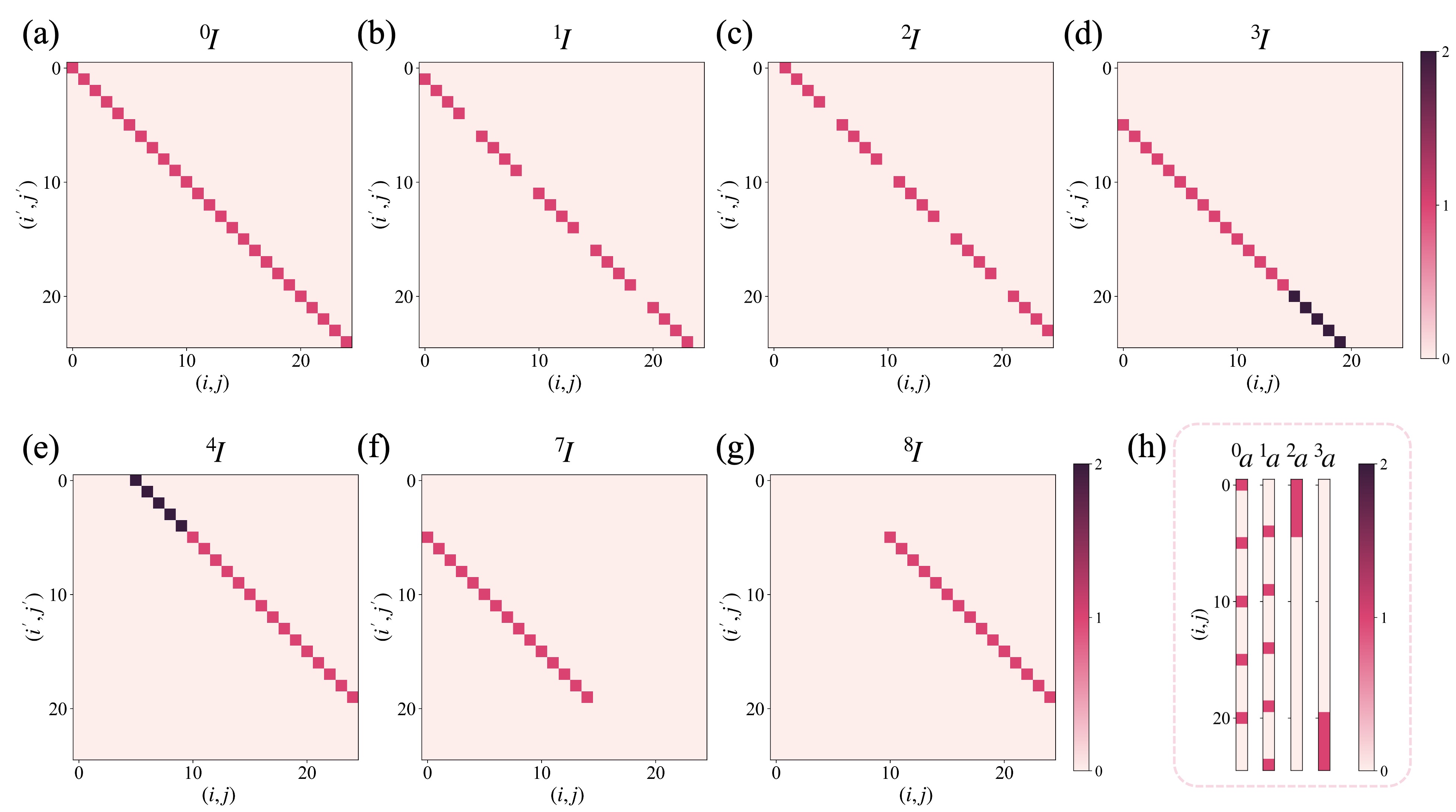}
			\caption{Visulizations of predefined matrices $I$ (a)-(g) and of predefined vectors $a$, for $N_x=N_z=5$. }
			\label{fig:Visualization_I_a}
		\end{figure}

		It is noticed that these matrices and vectors contain changeable values including $\alpha$, $\beta$, $\phi_{\mathrm{L}}$, $\phi_{\mathrm{R}}$, $\phi_{\mathrm{U}}^\prime$ and $\phi_{\mathrm{D}}^\prime$, which adds complexities of constructing them on the coding level. An convenient approach is to predefine matrices and vectors of which the elements are purely numbers, and sum them up, with $\alpha$, $\beta$, $\phi_{\mathrm{L}}$, $\phi_{\mathrm{R}}$, $\phi_{\mathrm{U}}^\prime$ and $\phi_{\mathrm{D}}^\prime$ being wieghts. Based on Tables.~(\ref{table:elements_A_b}) and (\ref{table:elements_B_c}), the predefined purely numbered matrices $I$ and vectors $a$ are derived and are listed in Table.~(\ref{table:elements_I_a}). It can be noticed that, six identical relations exist:
		\begin{equation}
			{}^5\!I^{i,j-1}_{i,j}={}^1\!I^{i,j-1}_{i,j},\quad {}^6\!I^{i,j+1}_{i,j}={}^2\!I^{i,j-1}_{i,j},\quad {}^4\!a_{i,j}={}^0\!a_{i,j},\quad {}^5\!a_{i,j}={}^1\!a_{i,j},\quad {}^6\!a_{i,j}={}^2\!a_{i,j},\quad {}^7\!a_{i,j}={}^3\!a_{i,j},
		\end{equation}
		\begin{table}[]
			\centering
			\renewcommand{\arraystretch}{1.5}
			\setlength{\tabcolsep}{2pt}
			\begin{tabular}{|c|c|c|c|c|c|c|c|c|c|c|c|}
				\hline
				& ${}^0\!I^{i,j}_{i,j}$ & ${}^1\!I^{i,j-1}_{i,j}$ & ${}^2\!I^{i,j+1}_{i,j}$ &${}^3\!I^{i-1,j}_{i,j}$ &${}^4\!I^{i+1,j}_{i,j}$ &${}^7\!I^{i-1,j}_{i,j}$&${}^8\!I^{i+1,j}_{i,j}$&${}^0\!a_{i,j}$&${}^1\!a_{i,j}$&${}^2\!a_{i,j}$&${}^3\!a_{i,j}$ \\ \hline
				Inner & 1 & 1 & 1 & 1 & 1 & 1 & 1 & 0 & 0 & 0 & 0 \\ \hline
				L     & 1 & / & 1 & 1 & 1 & 1 & 1 & 1 & 0 & 0 & 0 \\ \hline
				R     & 1 & 1 & / & 1 & 1 & 1 & 1 & 0 & 1 & 0 & 0 \\ \hline
				U     & 1 & 1 & 1 & / & 2 & / & 0 & 0 & 0 & 1 & 0 \\ \hline
				D     & 1 & 1 & 1 & 2 & / & 0 & / & 0 & 0 & 0 & 1 \\ \hline
				LU    & 1 & / & 1 & / & 2 & / & 0 & 1 & 0 & 1 & 0 \\ \hline
				LD    & 1 & / & 1 & 2 & / & 0 & / & 1 & 0 & 0 & 1 \\ \hline
				RU    & 1 & 1 & / & / & 2 & / & 0 & 0 & 1 & 1 & 0 \\ \hline
				RD    & 1 & 1 & / & 2 & / & 0 & / & 0 & 1 & 0 & 1 \\ \hline
			\end{tabular}
			\caption{Refined set of matrices $I$ and vectors $a$.}
			\label{table:elements_I_a_refined}
		\end{table}
		and the refined set is shown in Table.~(\ref{table:elements_I_a_refined}). The predefined matrices $I$ and vectors $a$ are visulized in Fig.~(\ref{fig:Visualization_I_a}), for $N_x=N_z=5$. And to this stage, we are finally able to solve the problem raised at the start of this subsection of finding the discrete forms of $\bar{\nabla}^2\phi$, $\partial_{\bar{x}}\phi$ and $\partial_{\bar{z}}\phi$, given $\alpha$, $\beta$, $N_x$, $N_z$ $\phi_\mathrm{L}$, $\phi_\mathrm{R}$, $\phi_\mathrm{U}^\prime$ and $\phi_\mathrm{D}^\prime$.
		\begin{eqnarray}
			\nonumber
			\left(\bar{\nabla}^2\phi\right)_{i,j}&=&\sum_{i^\prime\!,j^\prime}\Big[-2(\alpha^2+\beta^2)\cdot{}^0\!I+\alpha^2\cdot{}^1\!I+\alpha^2\cdot{}^2\!I+\beta^2\cdot{}^3\!I+\beta^2\cdot{}^4\!I\Big]_{i,j}^{i^\prime\!,j^\prime}\cdot\phi_{i^\prime\!,j^\prime}\\
			&+&\alpha^2\cdot\left(\phi_{\mathrm{L}}\right)_{i}\cdot{}^0\!a_{i,j}+\alpha^2\cdot\left(\phi_{\mathrm{R}}\right)_{i}\cdot{}^1\!a_{i,j}+2\beta\cdot\left(\phi_{\mathrm{U}}^\prime\right)_{j}\cdot{}^2\!a_{i,j}-2\beta\cdot\left(\phi^\prime_{\mathrm{D}}\right)_{j}\cdot{}^3\!a_{i,j}\\
			\left(\partial_{\bar{x}}\phi\right)_{i,j}&=&\sum_{i^\prime\!,j^\prime}\Big[-\frac{1}{2}\alpha\cdot{}^1\!I+\frac{1}{2}\alpha\cdot{}^2\!I\Big]_{i,j}^{i^\prime\!,j^\prime}\cdot\phi_{i^\prime\!,j^\prime}-\frac{1}{2}\alpha\cdot\left(\phi_{\mathrm{L}}\right)_{i}\cdot{}^0\!a_{i,j}+\frac{1}{2}\alpha\cdot\left(\phi_{\mathrm{R}}\right)_{i}\cdot{}^1\!a_{i,j}\\
			\left(\partial_{\bar{z}}\phi\right)_{i,j}&=&\sum_{i^\prime\!,j^\prime}\Big[\frac{1}{2}\beta\cdot{}^7\!I-\frac{1}{2}\beta\cdot{}^8\!I\Big]_{i,j}^{i^\prime\!,j^\prime}\cdot\phi_{i^\prime\!,j^\prime}+\left(\phi^\prime_{\mathrm{U}}\right)_{j}\cdot{}^2\!a_{i,j}+\left(\phi^\prime_{\mathrm{D}}\right)_{j}\cdot{}^3\!a_{i,j}		
		\end{eqnarray}
	
	\SIsubsection{2.4 Adaptive time-stepping strategy}
	
		To balance accuracy and computational efficiency, we adopt an adaptive time-stepping strategy. At each time level, the explicit Euler update is written as
		\begin{equation}
			(\bar{n}_\alpha)_{i,j,k+1}=(\bar{n}_\alpha)_{i,j,k}+		(\partial_{\bar{t}}\bar{n}_\alpha)_{i,j.k}\Delta \bar{t}
		\end{equation}
		% Adaptive time-stepping with Euler iteration
		where $\Delta \bar{t}$ is no longer fixed but determined dynamically based on the local variation of the solution. The adaptivity criterion is governed by a user-prescribed relative tolerance $\varepsilon$ (e.g., $\varepsilon = 0.01$). For each grid cell $(i,j)$, we compute the relative change of the primary unknown over the step:
		\begin{equation}
			R_{i,j} = \frac{\left|(\partial_{\bar{t}}\bar{n}_\alpha)_{i,j.k}\right|}
			{\max\left(\left|(\bar{n}_\alpha)_{i,j,k}\right|, \eta\right)},
		\end{equation}
		where $\eta$ is a small positive constant to avoid division by zero in regions of near-zero concentration. The step is accepted if the maximum relative change over all cells satisfies
		\begin{equation}
			\max_{i,j} R_{i,j} \le \varepsilon.
		\end{equation}
		If this condition is violated, the step is rejected and retried with a reduced step size (e.g., halved). Conversely, if the change is much smaller than $\varepsilon$, the subsequent step size is increased to improve efficiency. This feedback loop enables the time step to automatically adapt to the temporal stiffness and dynamics of the solution.
		
		The advantages of this adaptive time-stepping approach are as follows:
		\begin{itemize}
			\item \textbf{Controlled accuracy}: The relative error is bounded by the prescribed tolerance at each step, preventing excessive numerical errors during rapid transients.
			\item \textbf{Computational efficiency}: Larger steps are automatically taken during smooth or slowly varying periods, significantly reducing the total number of steps and CPU time without compromising accuracy.
			\item \textbf{Robustness}: The automatic rejection and retrial mechanism prevents instability or divergence caused by overly large steps, especially in stiff or highly nonlinear regimes.
			\item \textbf{Ease of use}: No tedious manual tuning of a fixed $\Delta t$ is required; the solver self-adjusts, making it practical for problems with widely varying temporal scales.
			\item \textbf{Optimal resource allocation}: Computational effort is concentrated where it is most needed (fast-changing regions) and saved elsewhere, achieving a favourable trade-off between accuracy and cost.
		\end{itemize}
		
\begin{figure*}[htbp!!!!!]
			\includegraphics[width=0.88\textwidth]{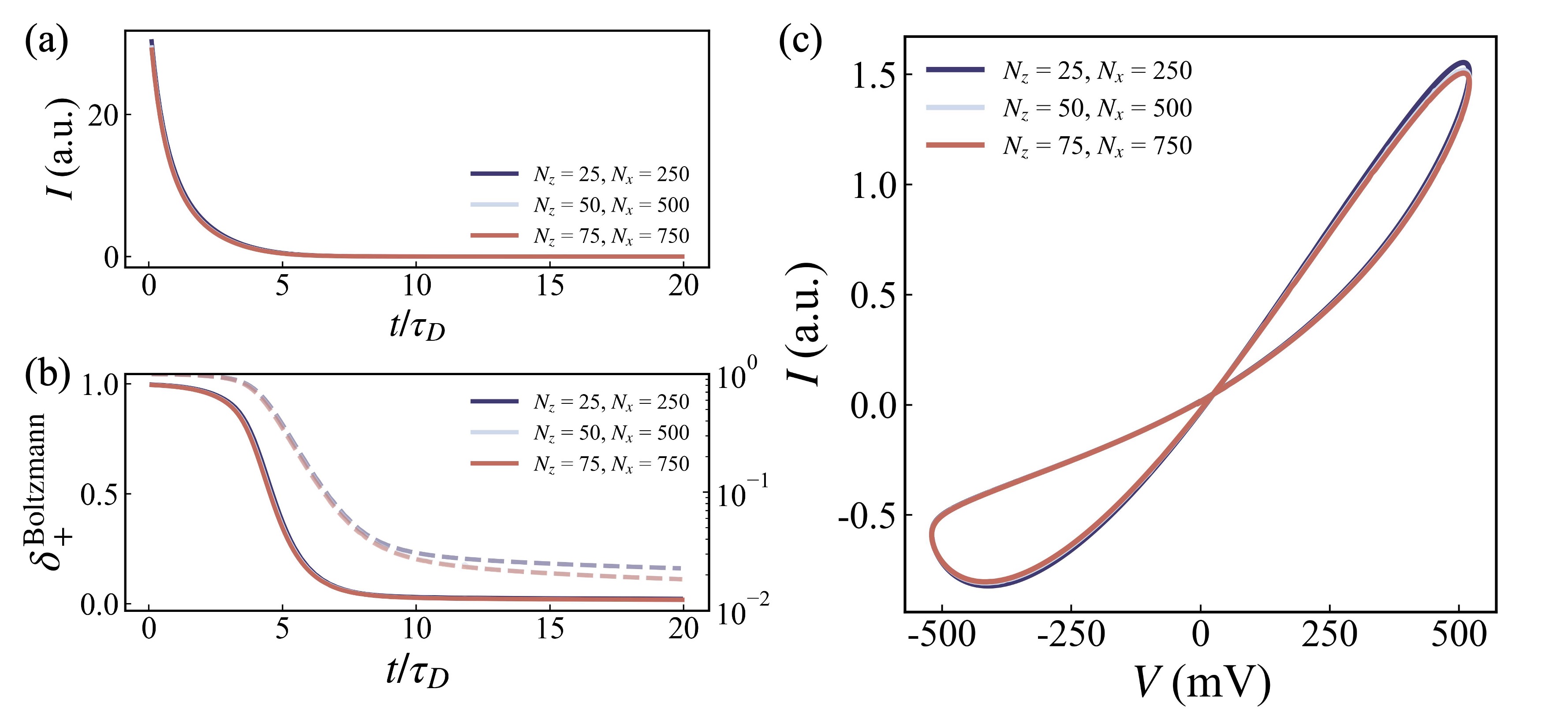}% Here is how to import EPS art
			\caption{(a) Benchmark against equilibrium state. (b) I-V loops at different grid densities.}
			\label{fig:Figure S5}
		\end{figure*}

		Therefore, the tolerance-based adaptive Euler iteration provides a simple yet effective foundation for time integration in our PDE solver.
	\clearpage
\SIsection{3. Simulated results}
	\SIsubsection{3.1 Benchmark of the simulations}
		We first benchmark the numerical solver by examining its numerical robustness and consistency with the known equilibrium state. When the surface charge is switched on at $t=0$ under zero applied voltage, ions drift from the reservoir into the nanochannel under the attraction of the charged surfaces (first enrichment), leading to a transient ionic enrichment on a characteristic timescale of $\tau_D$. As shown in Fig. \ref{fig:Figure S5}(a), the temporal evolution of the ionic current is essentially unchanged upon increasing the spatial grid density, demonstrating the numerical robustness of the solver against grid discretization. At long times, the system approaches equilibrium, where the ionic concentration is expected to follow the Boltzmann distribution. We therefore quantify the deviation of the simulated concentration profile from the Boltzmann distribution in Fig. \ref{fig:Figure S5}(b).
		\begin{equation}
			\delta_+^\mathrm{Boltzmann}=\frac{1}{Lh}\iint\left|\frac{n_+(x,z)-n_0\exp(-\Phi(x,z))}{n_+(x,z)+n_0\exp(-\Phi(x,z))}\right|\mathrm{d}x\mathrm{d}z
		\end{equation}
		 The relative deviation decreases toward zero as the system relaxes, confirming the consistency of the numerical evolution with the known equilibrium state. We further examine the non-equilibrium response under an applied time-varying voltage. As shown in Fig. \ref{fig:Figure S5}(c), the simulated I-V characteristics remain nearly unchanged for different grid densities, providing an additional test of grid convergence for the nonequilibrium solution.
		
\begin{figure*}[htbp!!!!!]
			\includegraphics[width=0.68\textwidth]{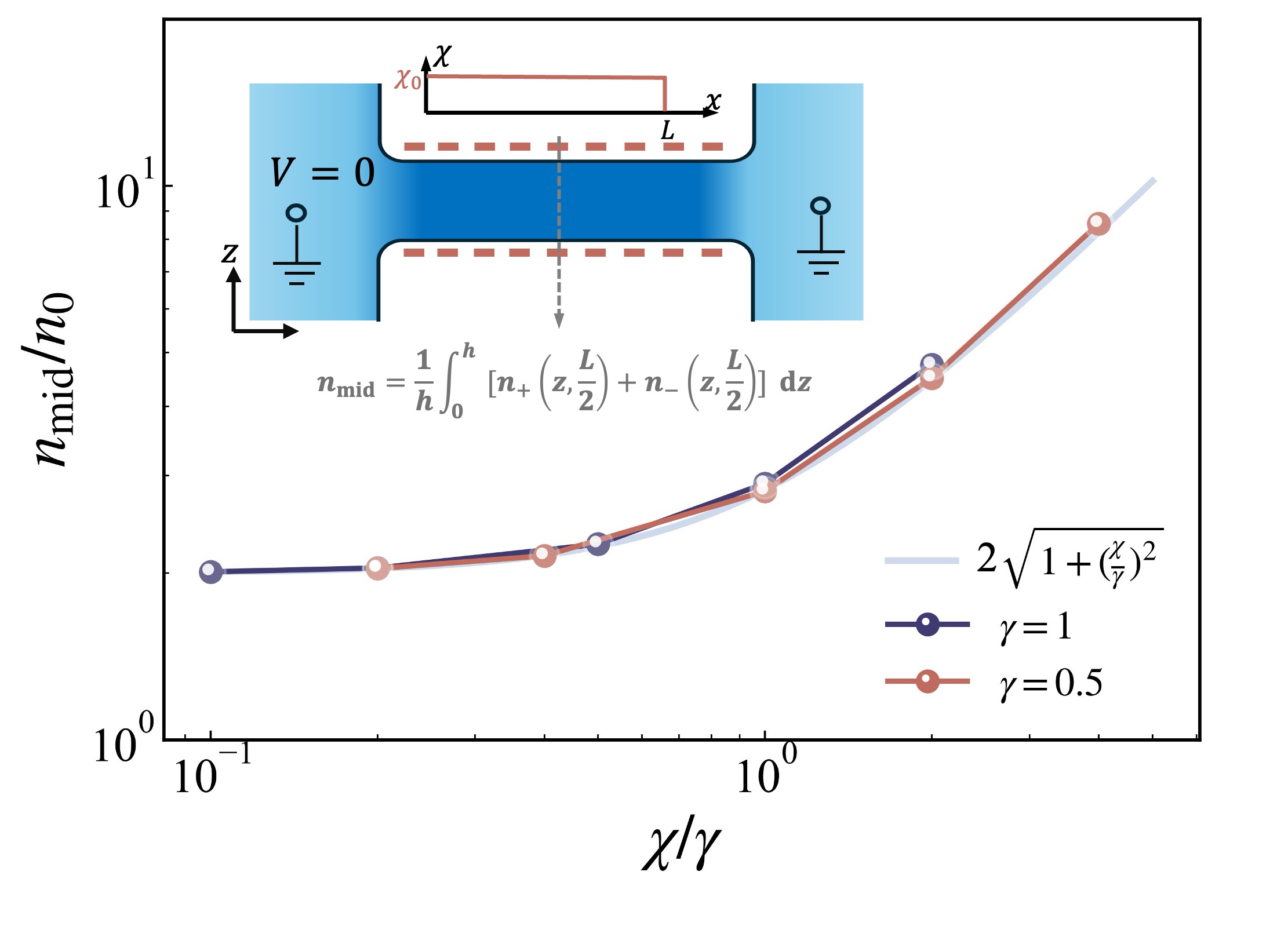}% Here is how to import EPS art
			\caption{Benchmark of the mid-channel ion concentrations.}
			\label{fig:Figure S6}
		\end{figure*}

		We further validate the solver against an analytical result. For a uniformly charged nanochannel approaching equilibrium, the simulated mean ion concentration at the channel center is compared with the corresponding analytical expression $n_\mathrm{mid}=2n_0\sqrt{1+(\frac{\chi}{\gamma})^2}$. As shown in Fig. \ref{fig:Figure S6}, the numerical results closely follow the analytical prediction over the investigated parameter range. 
		
		Together, the grid-convergence tests, recovery of the known equilibrium distribution, and agreement with the analytical solution establish the reliability of the numerical solver for both equilibrium and non-equilibrium ion transport. These tests assess both the numerical robustness and physical consistency of the solver, providing direct validation of the implemented PDE solving methods without relying on agreement with another numerical implementation such as COMSOL Multiphysics.
		
\begin{figure*}[htbp!!!!!]
			\includegraphics[width=0.85\textwidth]{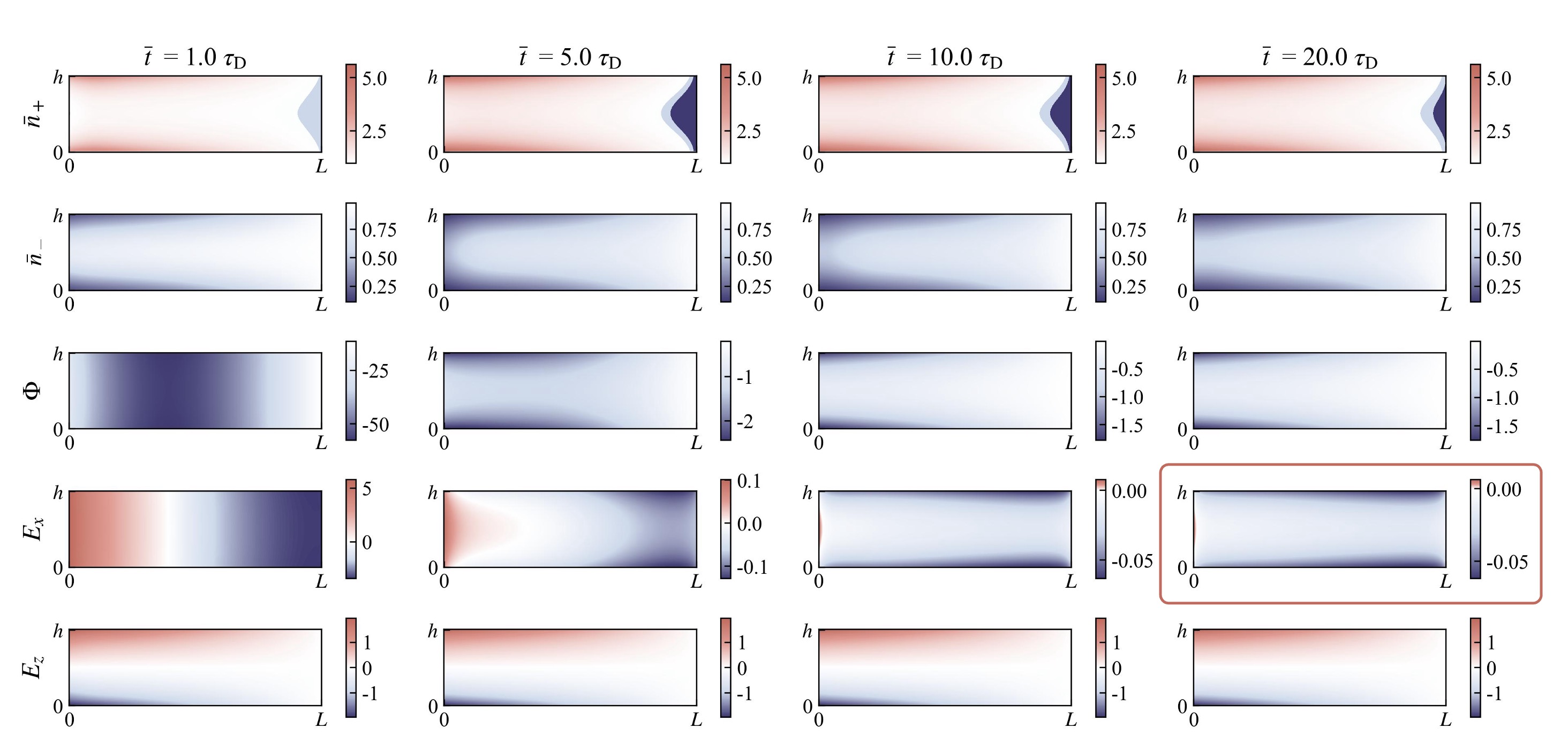}% Here is how to import EPS art
			\caption{The dynamics reaching equilibrium.}
			\label{fig:Figure S7}
		\end{figure*}

\begin{figure*}[htbp!!!!!]
			\includegraphics[width=0.85\textwidth]{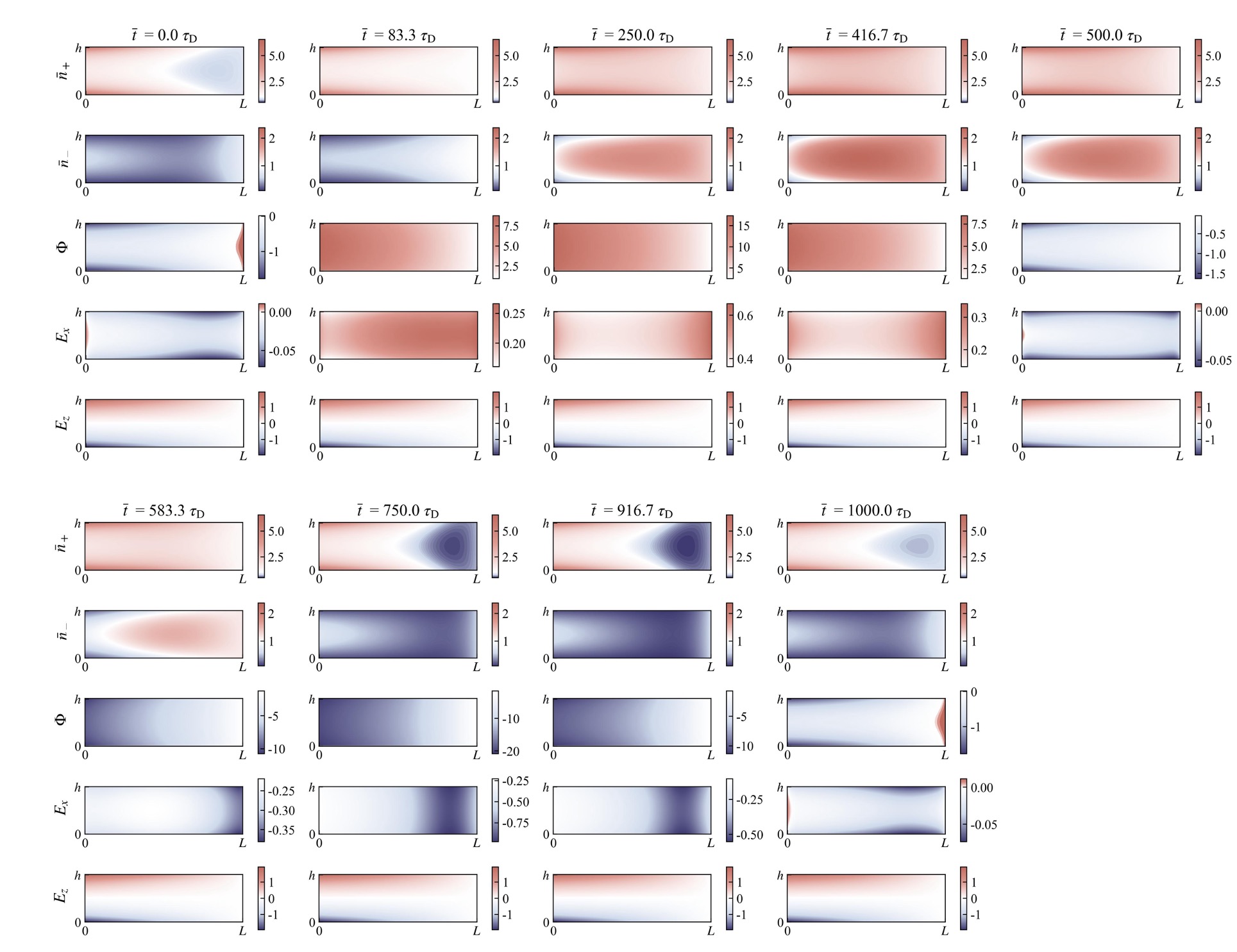}% Here is how to import EPS art
			\caption{The dynamics upon a sinusoidal voltage $V(t)=V_\mathrm{max}\sin(2\pi t/T)$ where $V_\mathrm{max} = $ 500 mV and $T=1000\tau_\mathrm{D}$.}
			\label{fig:Figure S8}
		\end{figure*}

		The detailed temporal evolution of ion concentration, electric potential and electric fields of the processes shown in Fig. \ref{fig:Figure S5}(a) and Fig. \ref{fig:Figure S5}(c) are shown in Fig. \ref{fig:Figure S7} and Fig. \ref{fig:Figure S8}, respectively.
		
		\clearpage
	\SIsubsection{3.2 The time-evolutions of various physical variables}
		
		\clearpage	
	\SIsubsection{3.3 The memristive characteristics explored over the parameter space. }
		\begin{figure*}[htbp!!!!!]
			\includegraphics[width=0.81\textwidth]{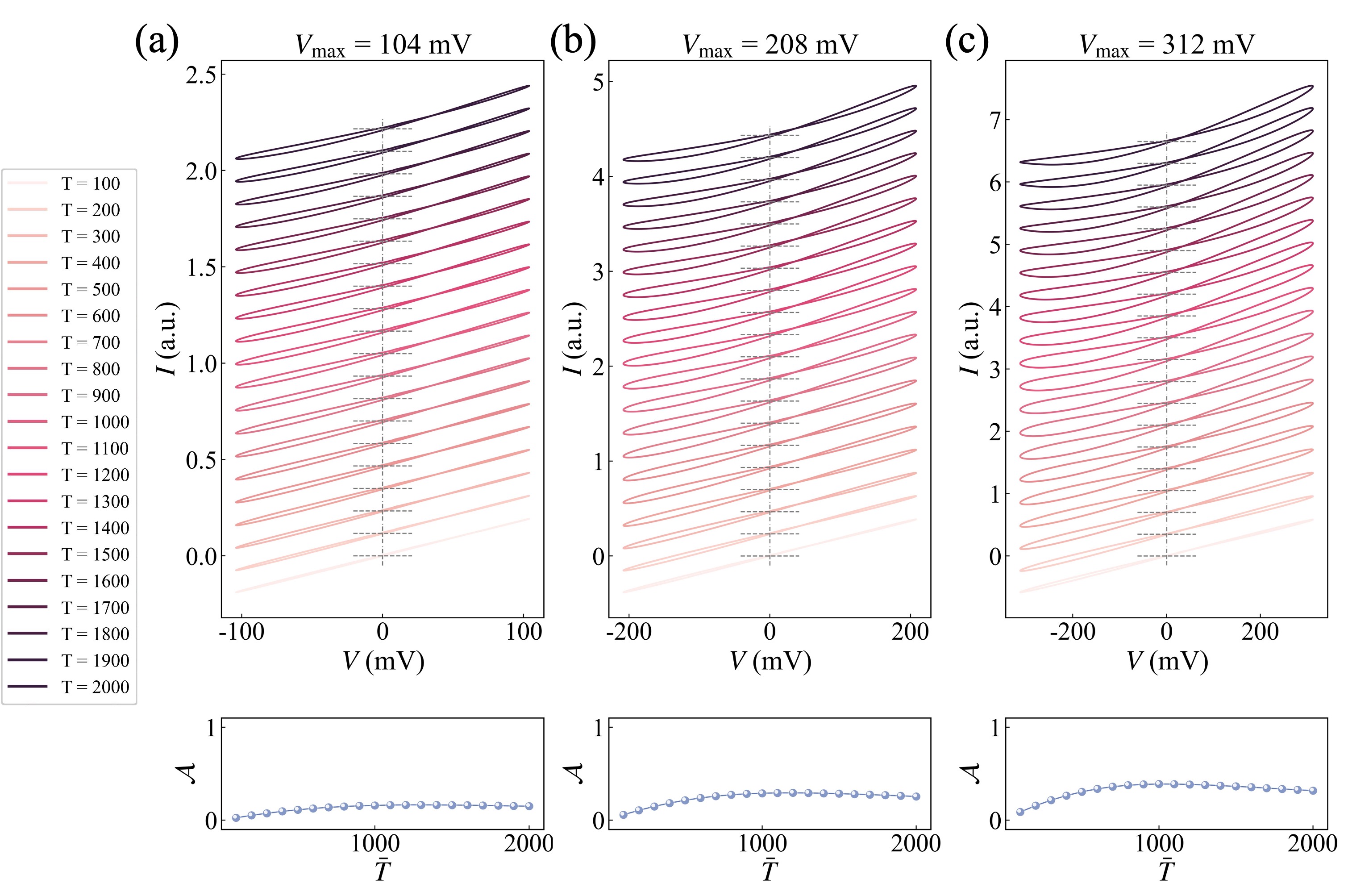}% Here is how to import EPS art
			\caption{The memristive characteristics at different $V_\mathrm{max}$ and $\gamma=\chi_0=1$.}
		\end{figure*}
		
		\begin{figure*}[htbp!!!!!]
			\includegraphics[width=0.85\textwidth]{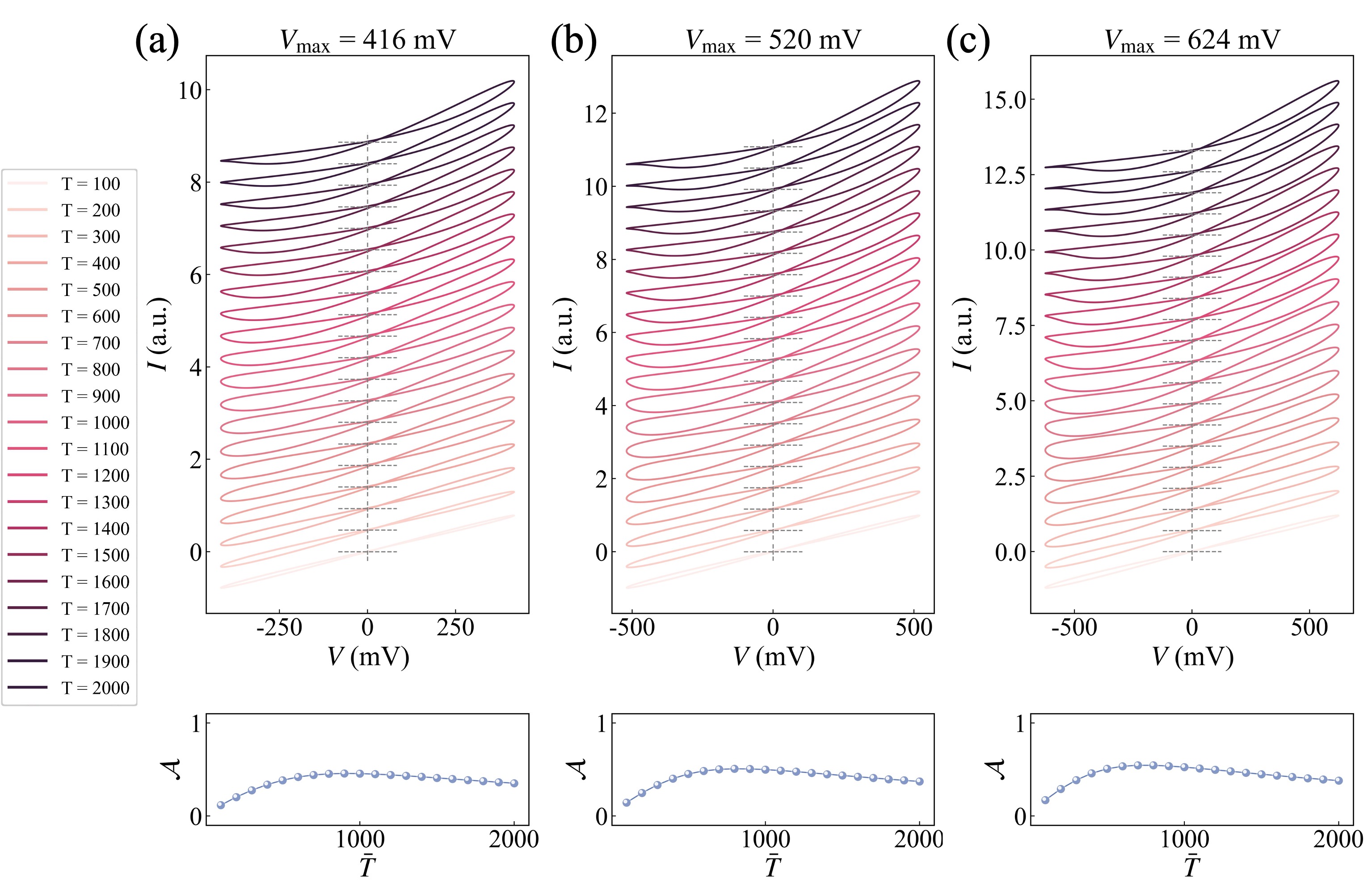}% Here is how to import EPS art
			\caption{The memristive characteristics at different $V_\mathrm{max}$ and $\gamma=\chi_0=1$.}
		\end{figure*}
		
		\begin{figure*}[htbp!!!!!]
			\includegraphics[width=0.85\textwidth]{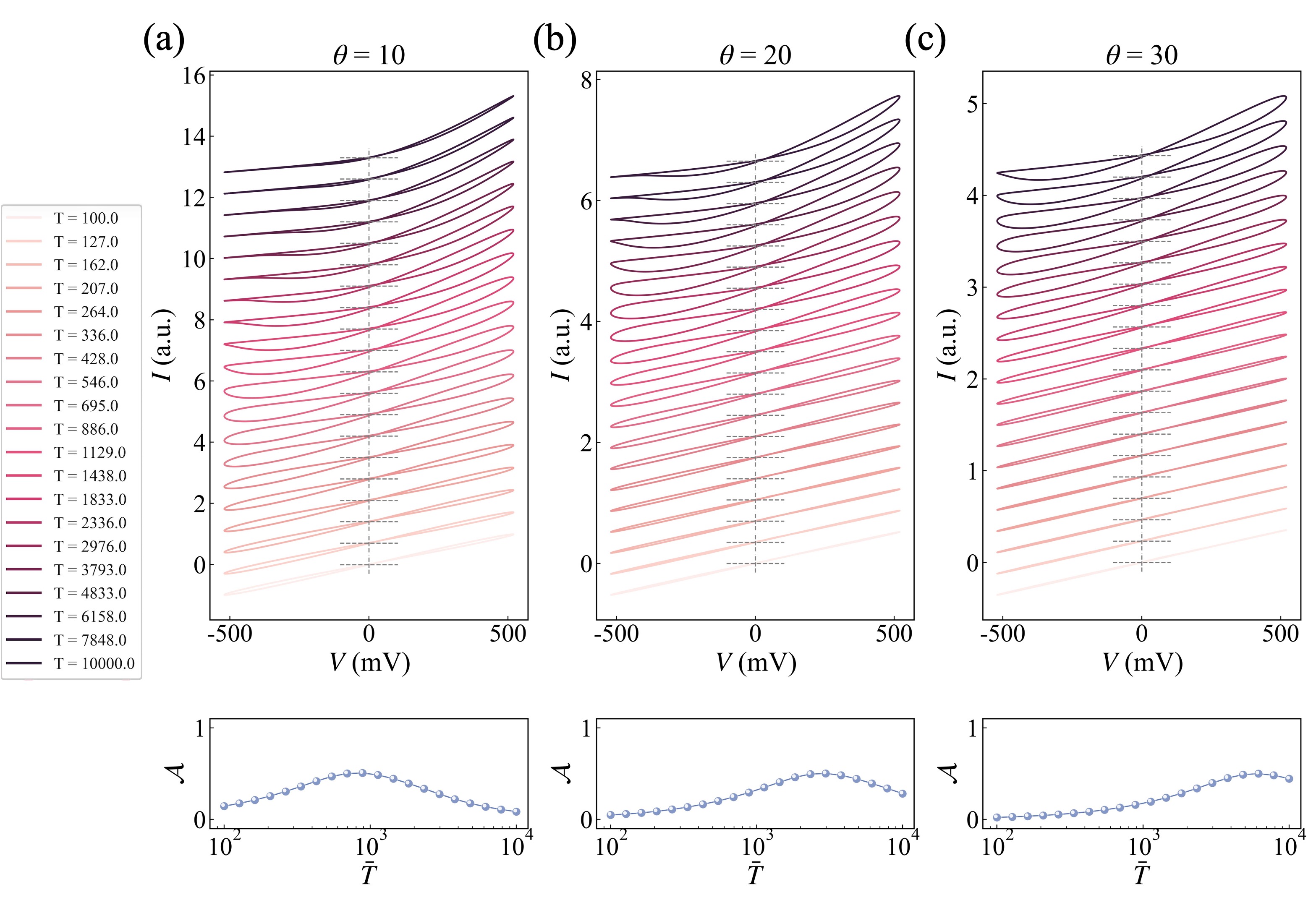}% Here is how to import EPS art
			\caption{The memristive characteristics at different $\theta$ and $\gamma=\chi_0=1$.}
		\end{figure*}
		
		\begin{figure*}[htbp!!!!!]
			\includegraphics[width=0.72\textwidth]{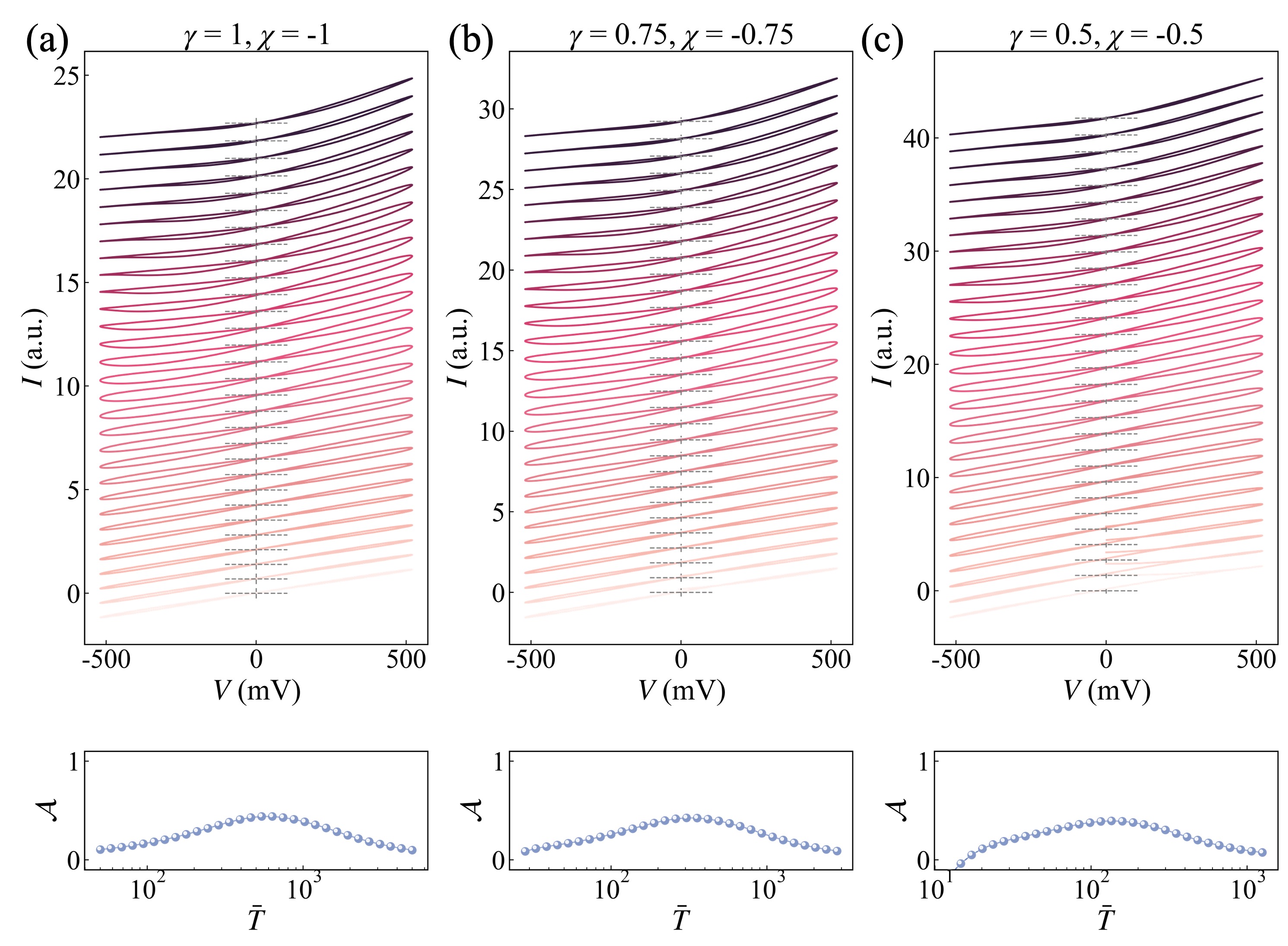}
			\caption{The memristive characteristics at different combinations of $\gamma$ and $\chi_0$.}
		\end{figure*}
		
		\begin{figure*}[htbp!!!!!]
			\includegraphics[width=0.88\textwidth]{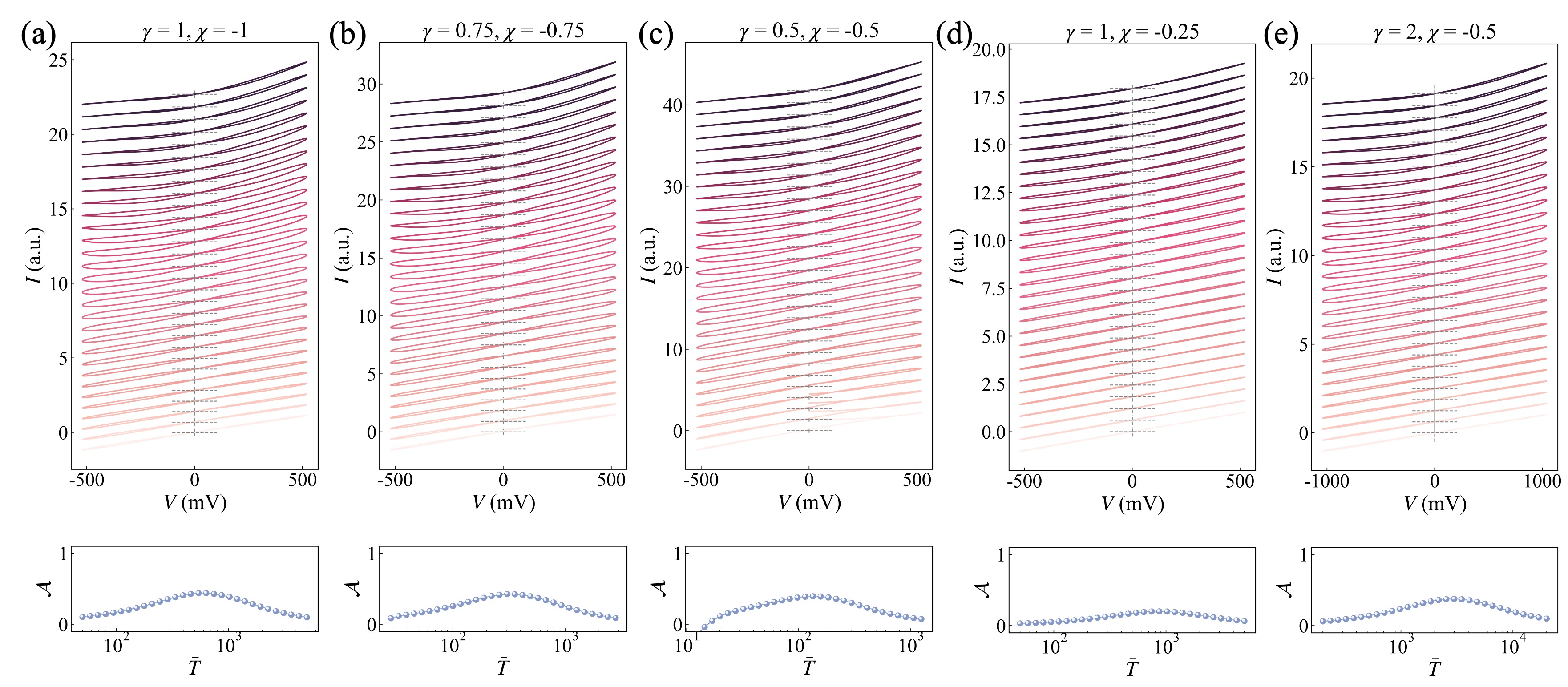}
			\caption{The memristive characteristics at different combinations of $\gamma$ and $\chi_0$.}
		\end{figure*}

\end{document}